\documentclass[twocolumn,prl,aps,superscriptaddress]{revtex4-1}
\usepackage[latin9]{inputenc}
\usepackage{mathtools}
\usepackage{amsbsy}
\usepackage{amstext}
\usepackage[unicode=true,pdfusetitle,
 bookmarks=false,
 breaklinks=false,pdfborder={0 0 1},backref=false,colorlinks=false]
 {hyperref}
\hypersetup{
 bookmarksnumbered=false,bookmarksopen=false}
\makeatletter

\@ifundefined{textcolor}{}
{%
 \definecolor{BLACK}{gray}{0}
 \definecolor{WHITE}{gray}{1}
 \definecolor{RED}{rgb}{1,0,0}
 \definecolor{GREEN}{rgb}{0,1,0}
 \definecolor{BLUE}{rgb}{0,0,1}
 \definecolor{CYAN}{cmyk}{1,0,0,0}
 \definecolor{MAGENTA}{cmyk}{0,1,0,0}
 \definecolor{YELLOW}{cmyk}{0,0,1,0}
}

\usepackage{amsmath}
\usepackage{graphicx}
\usepackage{amssymb}
\usepackage{txfonts,color}
\makeatother

\begin{document}

\title{Robust Oscillator-Mediated  Phase Gates Driven by Low-Intensity Pulses}

\author{I\~nigo Arrazola}
\email[email address: ]{iarrazola003@gmail.com} 
\affiliation{Vienna Center for Quantum Science and Technology, Atominstitut, TU Wien, 1040 Vienna, Austria}
\author{Jorge Casanova}
\affiliation{Department of Physical Chemistry, University of the Basque Country UPV/EHU, Apartado 644, 48080 Bilbao, Spain}
\affiliation{EHU Quantum Center, University of the Basque Country UPV/EHU, Leioa, Spain}
\affiliation{IKERBASQUE,  Basque  Foundation  for  Science, Plaza Euskadi 5, 48009 Bilbao,  Spain}

\begin{abstract}
\centerline{\large \bf Abstract}
\vspace{2mm}
Robust qubit-qubit interactions mediated by bosonic modes are central to many quantum technologies. Existing proposals combining fast oscillator-mediated gates with dynamical decoupling require strong pulses or fast control over the qubit-boson coupling. Here, we present a method based on dynamical decoupling techniques that leads to faster-than-dispersive entanglement gates with low-intensity pulses. Our method is general, i.e., it is applicable to any quantum platform that has qubits interacting with bosonic mediators via longitudinal coupling. Moreover, the protocol provides robustness to fluctuations in qubit frequencies and control fields, while also being resistant to common errors such as frequency shifts and heating in the mediator as well as crosstalk effects. We illustrate our method with an implementation for trapped ions coupled via  magnetic field gradients. With detailed numerical simulations, we show that entanglement gates with infidelities of $10^{-3}$ or $10^{-4}$ are possible with current or near-future experimental setups, respectively.
\end{abstract}
\maketitle

\section{Introduction} 
High-fidelity entanglement generation among qubits is crucial for quantum information processing~\cite{Nielsen}. In most platforms, entangling gates come via direct interactions (e.g. hyperfine fields among nuclear spins) or via a bosonic mediator. Examples of the latter include solid-state qubits coupled to microwave resonators~\cite{Jin12,Billangeon15,Richer16,Beaudoin16,Bosco22}, or trapped ions sharing vibrational modes~\cite{Mintert01}. 
In this scenario, the paradigmatic M{\o}lmer-S{\o}rensen (MS) gate~\cite{Sorensen99,Solano99,Sorensen00,Sackett00} and related schemes~\cite{Milburn00,Leibfried03,Zheng04} reach entanglement operations with inherent robustness to uncertainties in the bosonic state.

In recent years, several gate schemes have been developed that operate on the same MS principle but are also robust against other sources of error~\cite{Valahu22_2}. 
Some examples are frequency or amplitude-modulated gates decoupling from mode decoherence~\cite{Cross15,Haddadfarshi16,Webb18,Shapira18,Zarantonello19,Sutherland20}, from spectator modes~\cite{Hayes12,Choi14,Green15,Leung18,Schafer18,Lu19,Milne20} or from deviations in the qubit-boson coupling strength~\cite{Royer17,Ge19,Burd21,Shapira22}. 
In another vein, dynamical decoupling (DD) is a well established paradigm to protect qubits against decoherence~\cite{Viola98,Ban98}. 
Some continuous DD techniques have been demonstrated to be suitable for quantum gate implementations~\cite{Timoney11,Tan13,Harty16,Weidt16,Guo18}, while pulsed DD methods achieve increased robustness employing suited sequences such as XY8~\cite{Carr54,Meiboom58,Gullion69,Souza12,Kabytayev14,Munuera20,Dong21,Ezzell22} or AXY~\cite{Casanova15,Casanova17,Arrazola18}. However, the use of pulsed DD to protect oscillator-mediated gates has been mostly limited to dispersive regimes~\cite{Piltz13,Qiu21,Barthel22,Morong22}, and to few spin-echo~\cite{Paik16,Ballance16,Bazavan22} or rotary-echo~\cite{Tan13} pulses. Note that the application of several $\pi$ pulses is desirable for efficient elimination of time-varying noise.

In this regard, Manovitz et al.~\cite{Manovitz17} have experimentally shown
that the MS gate can be combined with pulse sequences given the ability to tune and turn on-and-off the qubit-boson coupling as many times as the number of DD pulses introduced. Another possibility explored theoretically is to combine an always-on qubit-boson coupling with strong $\pi$ pulses~\cite{Rabl10,Arrazola18,Rosenfeld21}. Although this is possible in certain trapped-ion architectures, turning on-and-off the qubit-boson coupling may be not practical in other platforms. On the other hand, the use of strong $\pi$ pulses is experimentally challenging since high-power controls are needed, while these induce crosstalk and hinder the applicability in multimode scenarios.

In this article, we design a DD sequence with low-intensity $\pi$ pulses --named TQXY16-- that achieves faster-than-dispersive entangling gates using static (i.e. non-tunable) qubit-oscillator coupling. 
Importantly, our gates decouple from dephasing, pulse imperfections, and unwanted finite-pulse effects, leading to high-fidelity.
Furthermore, we demonstrate the versatility of our protocol by incorporating techniques that lead to additional resilience to decoherence on the bosonic mediator and potential crosstalk effects. Although our method is general, we exemplify its performance in radio-frequency controlled trapped ions demonstrating infidelities within the $10^{-3}$ threshold at state-of-the-art experimental conditions, and of $10^{-4}$ in near-future setups.

\section{Results} 

\subsection{Gate with instantaneous pulses}
We consider a system that comprises two qubits and a bosonic mode --with frequencies $\omega_1, \omega_2$ and $\nu$-- coupled via longitudinal coupling~\cite{Mintert01,Jin12,Billangeon15,Richer16,Beaudoin16,Bosco22} (here, and throughout the paper, $H$ is $H/\hbar$, meaning all
Hamiltonians are given in units of angular frequency),
\begin{equation}\label{Hzero}
H_0=\nu a^\dagger a + \eta \nu (a+a^\dagger)S_z.
\end{equation}
Here, $a^\dagger (a)$ is the creation (annihilation) operator of the bosonic mode, $S_{\mu}\equiv\sigma_1^{\mu}+\sigma_2^\mu$ with $\mu\in{x,y,z}$ are collective qubit operators, and $\eta \nu$  is the coupling strength.
Also, note that $H_0$ is written in a rotating frame with respect to (w.r.t) the qubit free-energy Hamiltonian $H_q=\sum_\mu\omega_\mu\sigma_\mu^z/2$.
We assume the usual experimental scenario $\eta\ll1$, thus we stay away from other paradigms that require stronger qubit-boson couplings~\cite{GarciaRipoll03,Duan04,Steane14,Bentley13,Sameti21}.
$H_0$ contains no driving fields, while in our method we drive the qubits for two main reasons: (i) Accelerate the gate by making the qubits rotate at a frequency close to the bosonic frequency $\nu$, and (ii) Protection of the gate from qubit noise of the form $\epsilon_j(t)\sigma_j^z/2$ leading to dephasing.
When driving the qubits, $H_0$ is completed with the term $H_d(t)=\sum_{\mu=x,y}\Omega_\mu(t)S_\mu/2$. In an interaction picture w.r.t $H_d + \nu a^\dagger a$ we get  
\begin{equation}\label{Hamil3}
H(t)=\eta \nu (ae^{-i\nu t}+a^\dagger e^{i\nu t}) \sum_{\mu=x,y,z}f_\mu(t)S_\mu,
\end{equation}
where $\sum_{\mu=x,y,z}f_\mu(t)S_\mu=U^\dagger_d(t)S_zU_d(t)$ with $U_d(t)=\mathcal{T}\exp{[-i\int_0^t H_d(t')dt']}$ being the time-ordered propagator. See supplementary note 1 for additional details.

\begin{figure}
\centering
\includegraphics[width=1\linewidth]{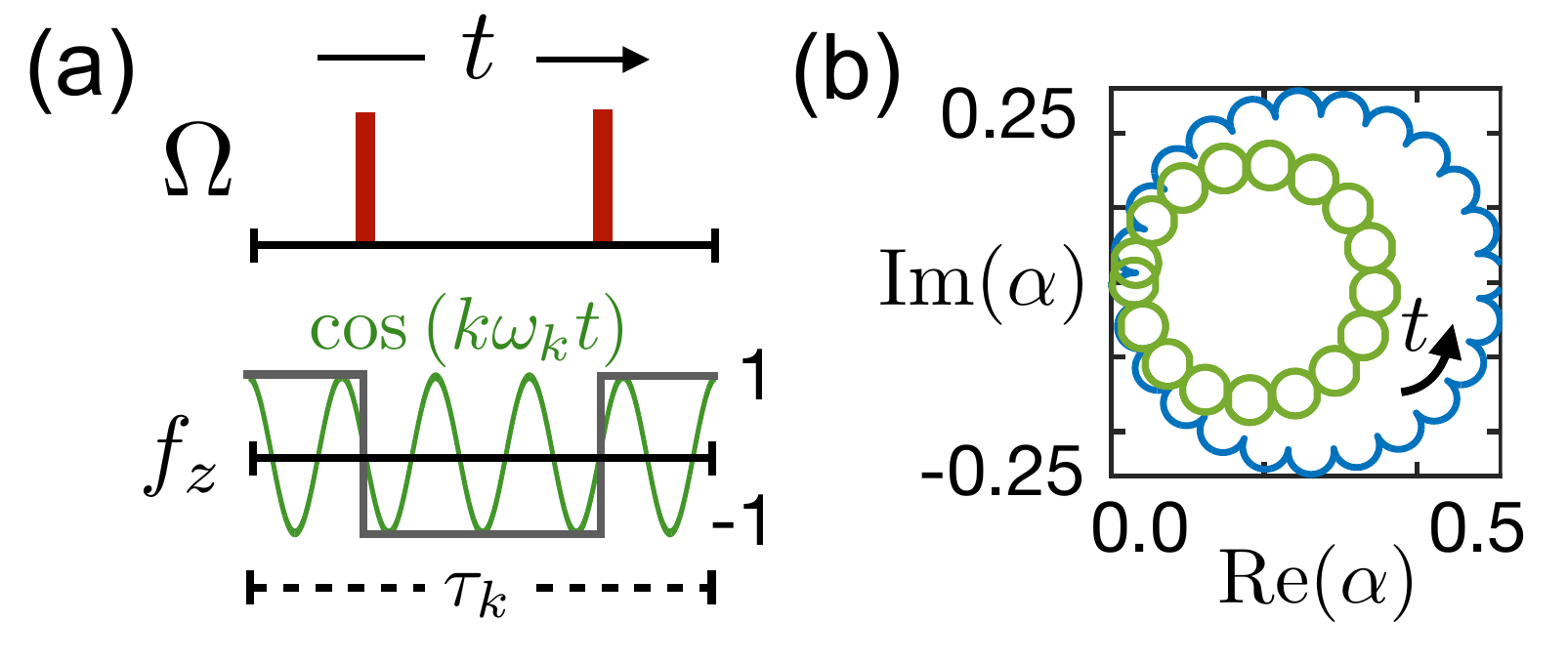}
\caption{Entangling gates with instantaneous $\pi$ pulses: (a)~Rabi frequency $\Omega(t)$ and modulation function $f_z(t)$ during a period $\tau_k$ for $k=5$. For comparison, we plot $\cos{(k\omega_k t)}$ in green. (b) Phase-space trajectory of $\alpha(t)$ during the application of a pulse sequence with $k=1$ and $k=5$ in blue and green, respectively.}\label{fig:1}
\end{figure}

If driving fields are delivered as instantaneous $\pi$ pulses (note this requires $\Omega_{x,y} \gg \nu$ during the application of the pulse) spaced $\tau/2$ apart, $f_{x,y}(t)$ can be neglected and $f_z(t) =1 (-1)$ if the number of applied pulses is even (odd), see the grey solid line in Fig.~\ref{fig:1}(a).
For the moment we consider instantaneous pulses, while later we treat the realistic case of non-instantaneous ones.
As  $\pi$ pulses are applied periodically, $f_z(t)$ takes the form of a function with period $\tau$ such that $f_z(t)=\sum_{n=1}^{\infty} f_n \cos{(n\omega t)}$, 
where $\omega=2\pi /\tau$ and $f_n=\frac{2}{\tau}\int_{0}^{\tau} dt' f_z(t')\cos{(n\omega t')}$.
Hence, under the assumption of instantaneous pulses we get
\begin{equation}\label{Hamil4}
H(t)=  \eta \nu\sum_{n=1}^{\infty} f_n  \cos{(n\omega t)} (ae^{-i\nu t}+a^\dagger e^{i\nu t})S_z,
\end{equation}
whilst setting an interpulse spacing $\tau/2=\tau_k/2$ such that  $\omega=\omega_k \approx \nu/k$ leads to a resonant qubit-boson interaction via the $k$th harmonic (from now on $\tau\rightarrow \tau_k$ and $\omega\rightarrow\omega_k$, where the subscript $k$ refers to the $k$th harmonic).
As $\eta\ll 1$, the terms in Eq.~(\ref{Hamil4}) that rotate with frequencies $\pm|\nu-n\omega_k|$ (where $n\neq k$) and $\pm|\nu+n\omega_k|$ can be substituted, using the rotating-wave approximation, by their second-order contribution (here, and in the rest of the paper, second-order stands for second order in $\eta$) leading to 
\begin{equation}\label{Hamil5}
H(t)\approx  \frac{1}{2}\eta \nu f_k (ae^{-i\xi_k t}+{\rm H.c.})S_z - \frac{1}{2}\eta^2 \nu J_kS_z^2,
\end{equation}
where $\xi_k= \nu -k\omega_k$ is the detuning w.r.t. the $k$th harmonic and $J_k= f_k^2/4 + \sum_{n\neq k} f_n^2/(1-n^2/k^2)$ is an effective spin-spin coupling constant that contains contributions from all harmonics. Note that, as $\eta\ll 1$ contributions of  higher order in $\eta$ can be neglected. See supplementary note 2 for additional details.
The propagator associated to Hamiltonian~(\ref{Hamil4}) is
\begin{equation}\label{Unitary1}
U(t)=\exp{\{[\alpha(t)a^\dagger-\alpha^*(t)a]S_z\}}\times \exp{[i\theta(t)S_z^2]}
\end{equation}
where $\alpha(t)=-i\eta \nu \int_0^t dt' f_z(t')e^{i\nu t'}\approx -\eta\nu f_k/(2\xi_k)(e^{i\xi_kt}-1)$ and

\begin{equation}\label{PhaseInt}
\theta(t)={\rm Im}\int_\mathcal{C} \alpha\, d\alpha \approx\frac{\eta^2\nu^2f_k^2}{4\xi_k}\Big[t-\frac{\sin{(\xi_k t)}}{\xi_k}\Big] + \frac{1}{2}\nu\eta^2J_k t,
\end{equation}
where $\mathcal{C}$ is the phase-space trajectory followed by $\alpha(t)$. Note that, if the gate time is chosen as $t_{\rm g}=2\pi/|\xi_k|$, $\alpha(t_{\rm g})\approx0$ at the end of the gate, making the gate insensitive to the bosonic state. To satisfy condition $\theta(t_{\rm g})=\pi/8$, we choose $\tau_k$ such that $\xi_k=2\eta \nu \Big\{ \sqrt{ f_k^2+4\eta^2J_k^2 } +2\eta J_k \Big\}$ for $J_k>0$. After a time $t_{\rm g}$ the propagator $U(t)$ approximates to $\exp{(i\frac{\pi}{8}S_z^2)}$. For two qubits, this is equivalent (up to a global qubit rotation) to the CPHASE gate, and transforms the state $|\!++\rangle$ into the Bell state $|\Phi^+\rangle=\frac{1}{\sqrt{2}}(|\!++\rangle+i|\!--\rangle)$. It is noteworthy that the choice of $\xi_k$ (thus $\tau_k$) is, in general, not trivial, as both $f_k$ and $J_k$ depend on $\tau_k$. However, in the cases discussed here this dependance does not hold, making the choice of $\xi_k$ direct. See supplementary note 3 for analytic expressions for $f_k$ and $J_k$. 

For instantaneous $\pi$ pulses one finds $f_k=f_k^{\rm ins}=\frac{4}{k\pi}\sin{(\frac{k\pi}{2})}$. Thus, if resonance is achieved via a low harmonic, e.g. $k=1$, the gate time is $t_{\rm g}^{k=1}\approx \pi^2/4\eta\nu$, a factor $\pi/2$ longer than the original MS gate. On the other hand, for sufficiently large harmonics the gate is mostly governed by the dispersive term $\frac{1}{2}\eta^2 \nu J_{k\rightarrow\infty}S_z^2\approx\eta^2 \nu S_z^2$ in Eq.~(\ref{Hamil5}), leading to $t_{\rm g}^{\rm d}=\pi/8\eta^2\nu$. We define {faster-than-dispersive} gates as those that satisfy $t^k_{\rm g}/t_{\rm g}^{\rm d}<1$.
For example, in the case $\eta=0.01$ and $k=1$ we find a faster-than-dispersive gate with $t_{\rm g}^{k=1}/t_{\rm g}^{\rm d}\approx 1/16$. This is, the gate is $16$ times faster than the dispersive one. In Fig.~\ref{fig:1}(b), we show $\alpha(t)$ for $\eta=0.03$, with $k=1$ and $5$.

It is noteworthy that the analysis conducted above is valid for $N_q$ qubits homogeneously coupled to the same bosonic mode, i.e. $S_\mu\rightarrow \sum_{j=1}^{N_q}\sigma_j^{\mu}$. For two qubits with inhomogeneous coupling, i.e. $S_z\rightarrow S_z=(\eta_1\sigma_1^z+\eta_2\sigma_2^z)/\eta$ where $\eta_j\ll1$, the method also yields to a CPHASE gate, however, in this case, the correct expression for the detuning $\xi_k$ is that in which every $\eta$ is substituted by $\sqrt{\eta_1\eta_2}$. For larger $\eta$, the rotating-wave approximation is not justified and terms neglected from Eq.~(\ref{Hamil4}) to  Eq.~(\ref{Hamil5}) will lead to significant residual qubit-boson entanglement at the end of the gate. See supplementary note 4 for additional details.

\subsection{Gate with low-intensity pulses}

In what follows, we discuss the realistic case of non-instantaneous pulses. For standard top-hat pulses the Fourier coefficient $f_k$ that quantifies the strength of the qubit-boson interaction reads (see supplementary note 3 for the derivation)
\begin{equation}\label{FourierTP}
f^{\rm th}_k\approx\frac{f^{\rm ins}_k}{1-\nu^2/\Omega^2}\cos{\bigg(\frac{\pi\nu}{2\Omega}\bigg)}.
\end{equation}
Notice that for {low-intensity pulses} --defined as those holding $\Omega < \nu$-- the value of $f_k$ decays with $(\Omega/\nu)^2$. As a result, achieving faster-than-dispersive gates is no longer possible. Note that $f_k$ directly relates to the non-dispersive contribution in $\theta(t)$, and, through condition $\theta(t_{\rm g})=\pi/8$, to the gate time $t_{\rm g}$. 

To solve this problem and optimize the strength of the qubit-boson interaction, we propose to modulate the Rabi frequency during the execution of each $\pi$ pulse. 
Specifically, we pose the following {ansatz} for $f_z(t)$
\begin{equation}\label{Ansatz}
f_z(t)=\cos[{\pi (t-t_i)/t_\pi]} + \beta(t)\sin{[k\omega_k (t-t_m)]},
\end{equation}
where $t_\pi$ is the $\pi$ pulse duration, and $t_i$ and $t_m=t_i+t_\pi/2$ are the initial and central points of the pulse. Note that the Rabi frequency is then given by $\Omega(t)=-\frac{\partial f_z(t)}{\partial t}\times[1-f^2_z(t)]^{-1/2}$. For the envelope function $\beta(t)$, we propose
\begin{equation}\label{envelope}
\beta(t)=\frac{d}{\pi k b}\sin(\pi k/2)\bigg[{\rm erf}\bigg(\frac{t-t_l}{ct_\pi}\bigg)-{\rm erf}\bigg(\frac{t-t_r}{ct_\pi}\bigg)\bigg],
\end{equation}
where $t_r=t_m+bt_\pi$ and $t_l=t_m-bt_\pi$. The free parameters $b$ and $c$ serve to control the width of the envelope function $\beta(t)$, while $d$ is proportional to its amplitude. 
Suitable values for $b, c$ and $d$ for the first harmonics are shown in table~\ref{table:2} (Methods).

From now on, we assume $t_\pi=\tau_k/2$, i.e., the pulse extends over a whole period $\tau_k/2$, leading to solutions with the lowest intensities.
As a result of our pulse design with suitable $b$ and $c$, the value for the Fourier coefficient $f_k$ is given by $f^{\rm m}_k=-{4 d}/{\pi k}$ where $|d|$ can take values from $0$ to $|d_{\rm max}|>1$. See supplementary note 3 for the derivation. Since now $f_k$ depends on $d$, this serves to control the strength of the interaction, thus the duration of the gate $t_{\rm g}$. Also, $d$ relates to the amplitude of the pulse, thus to the maximum value of the Rabi frequency $\Omega_{\rm pp}$. Typically, we look for large values of $d$, bounded by $d_{\rm max}$ or by the experimentally available $\Omega_{\rm pp}$.

Now we describe the recipe to design faster-than-dispersive gates using low-intensity pulses. First we choose a value for the harmonic $k$. Larger $k$ allow for lower pulse intensities at the price of longer gates. Second, we use Eq.~(\ref{Ansatz}) to generate the modulation function $f_z(t)$ and the Rabi frequency $\Omega(t)$ for different values of $d$, and calculate both the gate time $t_{\rm g}$ and $\Omega_{\rm pp}={\rm max}[|\Omega(t)|]$. We note that the obtained $\Omega(t)$ can lead to pulses along arbitrary axes (e.g. X or Y). 
In particular, for reasons described later, we target gates formed by concatenating blocks of $16$ pulses.
For that, the gate time $t_{\rm g}$ must be $8N\tau_k$, where $N$ is an integer number. 
This translates into the condition $(\nu-\xi_k)/8k|\xi_k|\in \mathbb{N}$ (note $t_{\rm g}=8N\tau_k$, while $t_{\rm g}=2\pi/|\xi_k|$ and $\tau_k=2\pi k/(\nu-\xi_k)$).
The final step is to select the values of $d$ for which this last condition is satisfied. As a result, we obtain all possible gates within the harmonic $k$ as well as the corresponding values for $\Omega_{\rm pp}$.

\begin{figure}
\centering
\includegraphics[width=1\linewidth]{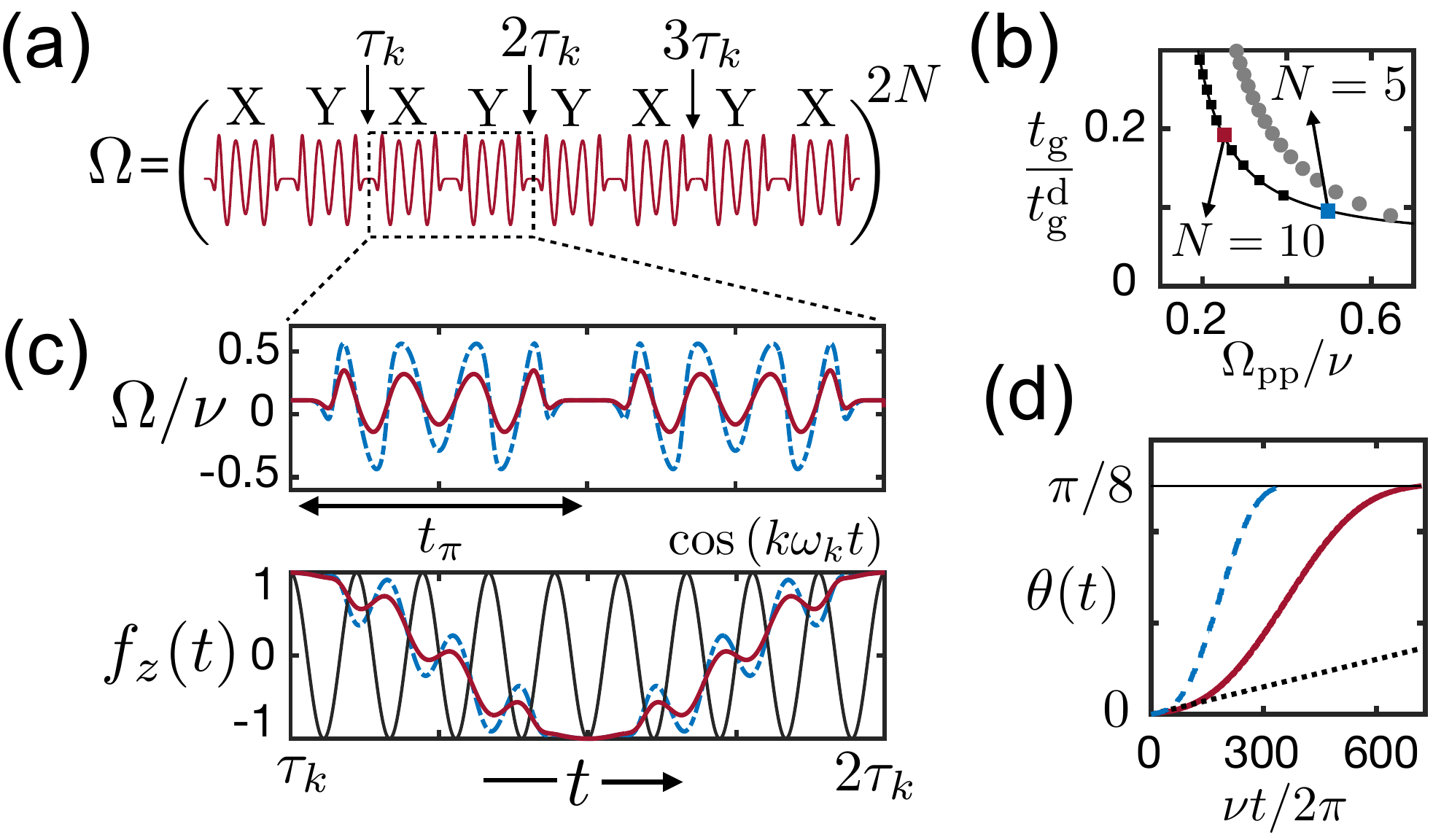}
\caption{ Amplitude modulated $\pi$ pulses: (a) $\Omega(t)$ for a single XY8 block. The whole sequence here is a concatenation of $2N$ blocks. (b) Gate time $t_{\rm g}$ as a function of $\Omega_{\rm pp}$, for values of $d$ between $0$ and $d_{\rm max}$. Values satisfying $t_{\rm g}= 8 N \tau_k$ are represented by square and round markers for $k=9$ and $k=7$, respectively. (c) $\Omega(t)$ and $f_z(t)$ for cases $N=5$ with $d=1.888$ (blue) and $N=10$ with $d=0.908$ (red) of the $9$th harmonic. (d) Gate phase $\theta(t)$ for $N=5$ (dashed blue line), $N=10$ (solid red line), and the dispersive case (dotted line). }\label{fig:2}
\end{figure}

Figure~\ref{fig:2}~(b) shows values of $t_{\rm g}$ and $\Omega_{\rm pp}$ obtained following the previous prescription
for $\eta=0.005$ and $k=7, 9$. 
Notice that there are plenty of solutions giving faster-than-dispersive gates, i.e. $t_{\rm g}/t_{\rm g}^{\rm d}<1$, using low-intensity pulses with values of $\Omega_{\rm pp}$ well below the frequency $\nu$.

As an example, we choose two solutions within the $9$th harmonic, where $\tau_k$ extends over approximately $9$ oscillator periods.
In Fig.~\ref{fig:2}~(c) the shapes of $\Omega(t)$ and $f_z(t)$ are displayed for cases $N=5$ and $10$.
Notice that $\Omega(t)$ achieves a larger amplitude when $N=5$.
As a consequence, it generates a faster gate. 
This is shown in Fig.~\ref{fig:2}~(d), where the two-qubit gate phase $\theta(t)$ related to the $N=5$ gate reaches the target value $\pi/8$ faster than the $N=10$ gate or the dispersive gate.

The reason for choosing the gate time as an integer multiple of $8\tau_k$ has to do with an efficient decoupling from finite-pulse effects produced by the terms $f_{x,y}(t)$ neglected in Eq.~(\ref{Hamil4}). 
In the same way, the XY8$\equiv$XYXYYXYX pulse structure assures cancelation of $\sigma^z$ type noise, as well as of Rabi frequency fluctuations.
In Fig.~\ref{fig:2}(a) the Rabi frequency is plotted ($\Omega_x(t)$ when ``X"; $\Omega_y(t)$ when ``Y") for an XY8 block.

To understand the elimination of finite-pulse effects, we calculate the second-order Hamiltonian of Eq.~(\ref{Hamil3}) after a XYXY block leading to (see supplementary note 5 for the derivation)
\begin{equation}
H_{\rm XYXY}=-\frac{1}{2}\eta^2\nu \Big\{ J_k^\perp (S_x^2 + S_y^2)   - B_k (a^\dagger a)S_z \Big\}.
\end{equation}
If our gate contains only XYXY blocks, $H_{\rm XYXY}$ adds to Hamiltonian~(\ref{Hamil5}) spoiling a high-fidelity performance.
To overcome this problem, we use a two-step strategy. Firstly, we concatenate XYXY and YXYX blocks (which form a XY8 block)
such that the term $B_k(a^\dagger a)S_z$ gets refocused. Note that in the presence of bosonic decoherence, this term will induce qubit dephasing.
Secondly, we cancel the remaining term $J_k^\perp (S_x^2 + S_y^2)$ by driving the two qubits with opposite phases every second XY8 block.
This is, when rotating by an angle $\pi$ the phase of the second qubit's driving $H_d^{(+)}=\sum_{\mu=x,y}\Omega_\mu(t)S^{(+)}_\mu/2$ becomes $H_d^{(-)}=\sum_{\mu=x,y}\Omega_\mu(t)S^{(-)}_\mu/2 $ instead, where $S_{\mu}^{(\pm)}\equiv\sigma_1^{\mu}\pm\sigma_2^{\mu}$.
This changes the sign of the $\sigma_1^{u}\sigma_2^{\mu}$ terms with $u, \mu \in \{x, y\}$, leading to refocusing of the term $J_k^\perp (S_x^2 + S_y^2)$ after every pair of XY8 blocks.  

Note that the second step requires the ability to address each qubit individually, and assumes that $N_q=2$, i.e. $S_\mu$ is given by the sum of two qubit operators. In the absence of individual addressing, one can incorporate the term $J_k^\perp (S_x^2 + S_y^2)$ into the gate, but then the operation applied is not equivalent to the CPHASE gate. For a discussion regarding this alternative gate, as well as the extension to the multiqubit case, see supplementary note 6.

Summarizing, our two-qubit gates are generated by nesting TQXY16$\equiv$XY8$^{(+)}$XY8$^{(-)}$ blocks, where TQXY16 stands for ``two-qubit" XY16, while XY8$^{(\pm)}$ imply qubits driven in phase or in anti-phase as discussed in the previous paragraph, while, importantly, each $\pi$ pulse is implemented according to the designs for $f_z(t)$ and  $\beta(t)$ presented in Eqs.~(\ref{Ansatz}, \ref{envelope}).

\subsection{Trapped-ion implementation \& Numerical results}
 We benchmark our method by simulating its performance in a pair of trapped ions in a static magnetic field gradient~\cite{Mintert01}.
In this scenario, qubit frequencies $\omega_\mu$ take values around $(2\pi)\times10$ GHz, $\nu=(2\pi)\times220$ kHz is the frequency of the centre-of-mass vibrational mode, $\eta=\gamma_e g_B/8\nu\sqrt{\hbar/M\nu}$ is an effective Lamb-Dicke factor where $\gamma_e=(2\pi)\times 2.8$ MHz/G, $g_B$ is the magnetic field gradient, and $M$ is the ion mass. The two-ion system has a second vibrational mode $``b"$ with its corresponding qubit-boson coupling. 
Thus, Hamiltonian~(\ref{Hzero}) is replaced by $H_0+H_{\rm 2M}$, where $H_{\rm 2M}=\sqrt{3}\nu b^\dagger b -3^{-1/4}\eta\nu(b+b^\dagger)S_z^{(-)}$.
The addition of $H_{\rm 2M}$ changes the dispersive coupling in Eq.~(\ref{Hamil5}) as $J_k\rightarrow J_k-1/3\sum_{n=1}^\infty f_n^2/(1-n^2/3k^2)$, which must be taken into account when following the prescription to calculate the valid gates. 
This step can be done for an arbitrary amount of spectator modes, given that the mode frequencies $\nu_m$ fulfil the condition $\eta f_n\ll |\nu_m-n\omega_k|$ for all odd $n$.

Although we simulate the performance of the gate with the two-mode Hamiltonian $H_f=H^{(\pm)}_d+H_0+H_{\rm 2M}$ (see column $\Delta\mathcal{I}_{2M}$ in table~\ref{table:1}), due to computational limitations we use the single-mode Hamiltonian $H_s=H_d^{(\pm)}+H_0+H_{\rm 2M}^{\rm eff}$ instead, where $H_{\rm 2M}^{\rm eff}=\frac{1}{3}\nu \eta^2 r S_z^2$ is the second-order contribution of $H_{\rm 2M}$. See supplementary note 7 for additional details. Here, $H_d^{(\pm)}$ stands for $H_d^{(+)}$ ($H_d^{(-)}$) every first (second) half of a TQXY16 block.

\begin{table*}[t!]
\caption{{Error budget}: Column XY8 and TQXY16 show the infidelities after an evolution with Hamiltonians $H_d^{(+)}+H_0+H_{\rm 2M}^{\rm eff}$  and $H_s$, respectively. The remaining columns show infidelities relative to the TQXY16 case (e.g. $\Delta\mathcal{I}_{\rm 2M}=\mathcal{I}_{\rm 2M}-\mathcal{I}_{\rm TQXY16}$), taking into account various experimental imperfections. In columns $\Delta\mathcal{I}_{\rm 2M}$, $\Delta\mathcal{I}_{\rm CT}$, and $\Delta\mathcal{I}_{\rm CT^*}$, infidelities obtained considering a second mode, crosstalk, and crosstalk with the sin$^2$ ramp are shown. In $\Delta\mathcal{I}_{ \delta\Omega}$, $\Delta\mathcal{I}_{\delta\nu}$, and $\Delta \mathcal{I}_{T_2}$ we show relative infidelities considering static shifts of $\delta\Omega=5\times10^{-3}$, $\delta\nu=10^{-5}$, and $\delta\omega=(2\pi)\times2\sqrt{2}$ kHz. $\dot{\bar{n}}$ shows the error considering heating with rates $\dot{\bar{n}}_1=35$ ph/s and $\dot{\bar{n}}_2=100$ ph/s for regimes (i) and (ii), respectively. The last column shows the overall error obtained by summing the values of all columns except those in $\mathcal{I}_{\rm XY8}$ and $\Delta\mathcal{I}_{\rm CT}$. }
\centering
\vspace{0.3cm}
\label{table:1}
\begin{tabular}{{ | c |c| c | c| c| c| c| c| c|  c| c|}}
\hline
 Gate & $\mathcal{I}_{\rm XY8}$ & $\mathcal{I}_{\rm TQXY16}$ & $\Delta\mathcal{I}_{\rm 2M}$ & $\Delta\mathcal{I}_{\rm CT}$ & $\Delta\mathcal{I}_{\rm CT*}$   & $ \Delta\mathcal{I}_{T_2}$ & $\Delta\mathcal{I}_{\delta\Omega}$ & $\Delta\mathcal{I}_{\delta\nu}$ & $\Delta\mathcal{I}_{\dot{\bar{n}}}$& $\mathcal{I}_{\rm total}$ ($10^{-4}$) \\
\hline 
G1  & $5.50$ & $0.01$    & $0.04$ & $24.8$  & $2.26$  &  $2.34$ &  $0.21$   & 0.28& $5.65$ & $10.8$ \\
\hline 
G2 & $28.7$ &  $<\!\!10^{-2}$    & $<\!10^{-2}$  & $3.20$  & $0.95$    & $2.45$ & $0.32$& $1.01$ &$19$ &  $23.7$  \\
\hline
G3  & $41.3$ & $<\!\!10^{-2}$    & $<\!10^{-2}$  & $134$  & $1.82$    &$2.35$ & $0.31$ &$0.26$ & $4.71$&  $9.45$\\
\hline
G4  & $>\!\!10^3$ & $0.12$    & $0.38$  & $0.23$  & $0.01$    & $0.43$ & $0.09$ &$<\!\!10^{-2}$& $0.11$ & $1.14$ \\
\hline
\end{tabular}
\end{table*}

We investigate two regimes: 
(i) $\eta=0.005$ ($g_B=19.16 \,\rm T/m$), which is the state-of-the-art of current experiments~\cite{Webb18,Barthel22}, and (ii) $\eta=0.04$ ($g_B=153.2\,\rm T/m$), which can be reached in near future setups~\cite{Weidt16}.

In regime (i), we consider three different gates, all within the $9$th harmonic. The first gate (G1), with a duration $t_{\rm g}= 1.64$ ms, appears after five TQXY16 blocks with pulse length $t_\pi=20.5 \ \mu$s  reaching $\Omega_{\rm pp}=(2\pi)\times124$~kHz. The second gate (G2) with gate time $t_{\rm g}=3.28$~ms uses ten TQXY16 blocks with pulse length $t_\pi=20.5 \ \mu$s reaching $\Omega_{\rm pp}=(2\pi)\times77.8$~kHz. 
The third gate (G3), with the gate-time $t_{\rm g}=3.94$ ms and $\Omega_{\rm pp}=(2\pi)\times78.69$ kHz, uses twelve blocks, each with a different pulse length and detuning, while it incorporates a technique to mitigate errors due to mode decoherence, see supplementary note 8.
In regime (ii) we consider a gate within the $5$th harmonic (G4). This gate occurs after two TQXY16 blocks where $t_{\rm g}=368~\mu$s,  $t_\pi=11.5~\mu$s, and $\Omega_{\rm pp}=(2\pi)\times80.9$ kHz. For further details regarding pulse parameters, see supplementary note 7.

The performance of the four gates in the presence of distinct error sources is shown in table~\ref{table:1}. Each simulated experiment starts from the state $|\!+_x+_y\rangle$ and targets the Bell-state $|\tilde{\Phi}^+\rangle=\frac{1}{\sqrt{2}}(|\!+_x+_y\rangle+i|\!-_x-_y\rangle)$, while in all cases we consider an initial motional thermal state with $\bar{n}=1$~\cite{Barthel22}. Other initial states result in similar values for the fidelity.

In the 2nd and 3rd columns of table~\ref{table:1} we show the gate error $\mathcal{I}=1-\mathcal{F}$ obtained by concatenating XY8 or TQXY16 blocks, respectively. Here, $\mathcal{F}=\langle \tilde{\Phi}^+| \rho|\tilde{\Phi}^+\rangle/\sqrt{{\rm Tr}(\rho^2)}$~\cite{Wang08}, where $\rho$ is the final state after tracing out the bosonic states. Notice that TQXY16 blocks achieve a clearly superior performance due to efficient decoupling from finite pulse effects. For these, $\mathcal{I}_{\rm TQXY16}\leq10^{-6}$ for all gates except G4, where finite the residual qubit-boson entanglement limits the error to approximately $10^{-5}$. In the fourth column we evaluate the effect of the second mode $b$ by numerically simulating the two-mode Hamiltonian $H_f$ (initialising the second mode $b$ in a thermal state with $\bar{n}=1$), which results in $\mathcal{I}_{\rm 2M}$.
The infidelities relative to the previous case (i.e. $\Delta\mathcal{I}_{\rm 2M}=\mathcal{I}_{\rm 2M}-\mathcal{I}_{\rm TQXY16}$) are given in the ``$\Delta\mathcal{I}_{\rm 2M}$" column of table~\ref{table:1}. 
Again, the effect of the second mode is relevant only for G4, which contributes $3.8\times10^{-5}$ to the total error. 
Importantly, this demonstrates that our gate is compatible with the presence of spectator modes.

To investigate the effect of crosstalk, we add the term $H^{(\pm)}_c=\sum_{\mu=x,y}\frac{\Omega_\mu(t)}{2}(\sigma_2^-e^{-i\Delta\omega t} \pm \sigma_1^-e^{i\Delta\omega t} +{\rm H.c.} )$ to $H_s$, where $\Delta\omega/(2\pi)=(\omega_2-\omega_1)/(2\pi)=2.54$ and $20.34$ MHz for regimes (i) and (ii), respectively.  The results are given in the ``$\Delta\mathcal{I}_{\rm CT}$" column of table~\ref{table:1}. 
In contrast to the effect of the spectator mode, crosstalk is most harmless with the G4 gate. This is expected, as G4 operates with a larger qubit detuning $\Delta \omega$ than the rest, while using a similar Rabi frequency.
To reduce the impact of crosstalk, we combine our pulses with sin$^2$-shaped ramps at the beginning and end of each pulse, see supplementary note 7, and optimize the length of the ramp using numerical simulations.  The resulting infidelities are shown in the  ``$\Delta\mathcal{I}_{\rm CT*}$"  column. Note that the sin$^2$ ramp reduces the value of $\mathcal{I}$ by at least an order of magnitude in most cases.

\begin{figure}
\centering
\includegraphics[width=1\linewidth]{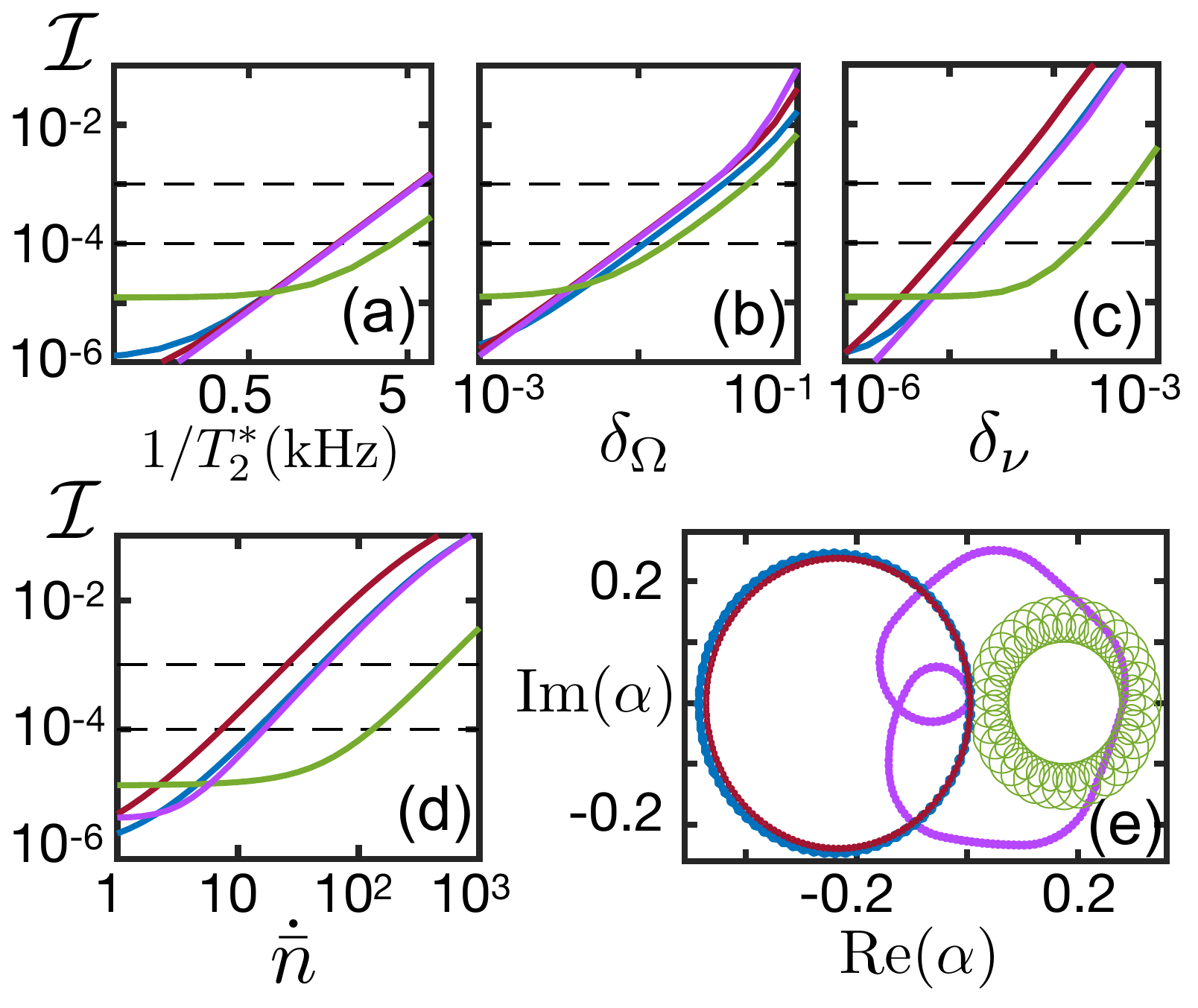}
\caption{ Sensitivity to errors: (a) Gate error $\mathcal{I}$ versus $1/T_2^*$ for the gates G1 (blue), G2 (red), G3 (purple) and G4 (green). In (b) and (c), the error is shown for Rabi frequency shifts $\Omega(t)\rightarrow(1\pm\delta_\Omega)\Omega(t)$ and in the mode frequency $\nu\rightarrow(1\pm\delta_\nu)\nu$. (d) Error under heating for different rates $\dot{\bar{n}}$. (e) Trajectory of $\alpha(t)$ for gates G1 (circular, blue), G2 (circular, red), G3  (non-circular, purple), and G4 (circular, green).}\label{fig:3}
\end{figure}

Robustness w.r.t.~common errors such as dephasing over qubits due to static shifts $\omega_j\rightarrow \omega_j \pm\delta\omega$, Rabi-frequency shifts (i.e. $\Omega(t)\rightarrow(1\pm\delta_\Omega)\,\Omega(t)$), and shifts on the mode frequency, $\nu\rightarrow(1\pm\delta_\nu)\,\nu$, is shown in Figs.~\ref{fig:3}~(a-c), where the infidelity is plotted versus the degree of uncertainty. In columns 7-9 of table~\ref{table:1}, we display the relative infidelities for a dephasing time $T_2^*\approx 500 \mu$s~\cite{Barthel22}, a Rabi-frequency shift $\delta_\Omega=5\times10^{-3}$, and a mode-frequency shift of $\delta_\nu=10^{-5}$~\cite{Johnson16}. For further details, see Methods. Furthermore, in Fig.~\ref{fig:3}~(d) we plot the infidelity versus $\dot{\bar{n}}$, while in column $\dot{\bar{n}}$, we show the relative infidelities for G1-3 and G4 for mode heating rates $\dot{\bar{n}}=35$ and $100$ ph/s, respectively. For further details, see supplementary note 7. Fig.~\ref{fig:3}~(e) shows the phase-space trajectory of $\alpha(t)$ for all gates G1-4.

Table~\ref{table:1} shows that mode heating is is the main source of error for gates in regime (i). This, along with dephasing and crosstalk, limits the fidelity of these gates to the $10^{-3}$ regime.
Despite its longer gate duration, G3 achieves better performance in terms of motion-induced errors than G1 and G2, proving the validity of the mode decoherence protecting technique described in the supplementary note 8. Finally, table~\ref{table:1} shows that G4 is the most robust w.r.t. experimental imperfections. This is reasonable since it uses a larger $\eta$ and is an order of magnitude faster than the other gates. In particular, we find that G4 achieves infidelities on the $10^{-4}$ regime, mainly limited by residual qubit-boson entanglement caused by off-resonant harmonics and the spectator mode. Note that the influence of this error has been taken into account in the supplementary note 3, where we also discuss potential effects of micromotion.

\section{Discussion} 

We have presented a DD sequence (TQXY16) based on the delivery of low-intensity $\pi$ pulses that achieve faster-than-dispersive two-qubit gates. Without the need of any numerical optimisation, we have designed entangling gates which are robust to fluctuations in qubit frequencies and control fields, as well as to finite-pulse effects hindering a high-fidelity performance. In addition, we have demonstrated the versatility of our protocol to adopt forms that provide an increased robustness against crosstalk and mode decoherence.

Our scheme is best suited for systems i) using longitudinal qubit-boson coupling with $\eta\ll1$, ii) where dephasing is the main source of qubit decoherence, and iii) where the Rabi frequencies $\Omega(t)$ are of the order (or far below) the mode frequencies $\nu$. This is the case, e.g., for spin qubits coupled to microwave cavities~\cite{Beaudoin16,Bosco22}. In Bosco et al.~\cite{Bosco22}, $\eta\sim10^{-2}$ and $\Omega/\nu\sim 0.1$. Superconducting qubit architectures exploiting longitudinal qubit-boson coupling have also been proposed~\cite{Billangeon15,Richer16}. Our method is also well suited for these systems when working with small $\eta$.

Finally, we tested the performance of our protocol in trapped ions coupled via static magnetic field gradients, where conditions (i), (ii), and (iii) are perfectly satisfied. Compared to existing multi-level schemes~\cite{Webb18,Weidt16}, our method has the advantage of using only two levels, which lowers the experimental requirements. Compared to previous pulsed DD methods~\cite{Arrazola18}, our method has the advantage of using realistic pulse intensities. Using detailed numerical simulations, we have obtained infidelities within the $10^{-3}$ threshold at state-of-the-art conditions, and in the $10^{-4}$ regime in near-future setups.

\begin{acknowledgements}
\section*{Acknowledgements}  I. A. would like to thank P. Rabl and J. S. Pedernales for useful discussions. I. A. acknowledges support from the European Union's Horizon2020 research and innovation programme under Grant Agreement No.~899354 (SuperQuLAN). J.~C. acknowledges the Ram\'{o}n y Cajal  (RYC2018-025197-I) research fellowship, the financial support from Spanish Government via EUR2020-112117 and Nanoscale NMR and complex systems (PID2021-126694NB-C21) projects, the EU FET Open Grant Quromorphic (828826), the ELKARTEK project Dispositivos en Tecnolog\'i{a}s Cu\'{a}nticas (KK-2022/00062), and the Basque Government grant IT1470-22.
\end{acknowledgements}

\section*{Methods} In table~\ref{table:2} we show suitable values of $b$ and $c$ given the harmonic $k$. Also, we show the maximum value of $|d|$ for which a physical pulse (i.e, $|f_z|(t)\leq 1$) can still be generated with the ansatz given in Eqs.~(\ref{Ansatz}, \ref{envelope}). For gates G1, G2 and G4, the selected values of $d$ are $1.915$, $0.933$ and $-0.321$, respectively. The list of detunings used in gate G3 is $\vec{\xi}_k=(2\pi)\times [$1.24, 0.31, 0.64, 0.09, 0.55, 0.05, 0.54, 0.06, 0.57, 0.14, 0.73, 0.80$]$ kHz. 

Our numerical simulations for dephasing consider an additional $\pm\delta\omega/2 S_z$ term in $H_s$, where $\delta\omega=\sqrt{2}/T_2^*$. In all three cases, each point is the average error obtained by a positive (e.g. $\omega_j\rightarrow \omega_j+\delta\omega$) and a negative (e.g. $\omega_j\rightarrow \omega_j-\delta\omega$) displacement. 

\begin{table}[ht]
\caption{Suitable pulse parameters. Suitable values the for $b$, $c$ and $|d_{\rm max}|$.  $|d_{\rm max}|$ corresponds to  the maximum value of $|d|$ for which a physical pulse (i.e, $|f_z|(t)\leq 1$) can still be generated.}
\centering
\vspace{0.3cm}
\label{table:2}
\begin{tabular}{{ |c| c | c| c| c| c| c| c| }}
\hline
 $k$ & $3$ & $5$ & $7$ & $9$ & $11$ & $13$ & $15$ \\
\hline
$b$ & $0.33$    & $0.30$ & $0.29$  & $0.33$ & $0.34$ & $0.35$ & $0.30$  \\
\hline
$c$ & $0.035$    & $0.04$  & $0.05$  & $0.042$ & $0.03$  & $0.035$ & $0.035$ \\
\hline
$d_{\rm max}$ & $-2.3$    & $-1.5$  & $2.4$  & $2.3$ & $1.9$  &$1.7$ & $2.3$   \\
\hline
\end{tabular}
\end{table}

\section*{Data availavility}
Data sharing not applicable to this article as no datasets were generated or analysed during the current study.

\section*{Author Contributions}
I. A. and J. C. conceived the idea. I. A. performed all the calculations. I. A. and J. C. wrote the manuscript.

\section*{Competing interests}
The authors declare no competing interests.

\clearpage

\widetext
\begin{center}
\textbf{ \large Supplemental Material:}
\textbf{ \large Robust Oscillator-Mediated  Phase Gates Driven by Low-Intensity Pulses }
\end{center}

\setcounter{equation}{0} \setcounter{figure}{0} \setcounter{table}{0}
\makeatletter \global\long\def\theequation{S\arabic{equation}}
 \global\long\def\thefigure{S\arabic{figure}}
 \global\long\def\bibnumfmt#1{[S#1]}
 \global\long\def\citenumfont#1{S#1}

 \section*{Supplementary note 1: The driving frame}
Here, we explain how to obtain Eq.~(2) from the main text, by going to the appropriate interaction picture. We start by adding the driving Hamiltonian $H_d(t)$ to $H_0$. Then, the total Hamiltonian reads
\begin{equation}
H_{I}(t)=\nu a^\dagger a +\frac{\Omega_x(t)}{2}S_x+\frac{\Omega_y(t)}{2}S_y+ \eta\nu(a+a^\dagger)S_z.
\end{equation}
In the interaction picture (labeled by {\it II}) with respect to $H_d(t)+\nu a^\dagger a$, the quantum state relates to the previous state as $\psi_{II}(t)=U^\dagger_{II}(t)\psi_I(t)$, where $U_{II}(t)$ is the time-evolution operator associated to $H_d(t)+\nu a^\dagger a$, that is $U_{II}(t)=e^{-i\nu a^\dagger a t}U_d(t)$ with $U_{d}(t)=\mathcal{T}\exp{[-i\int_0^t H_d(t')dt']}$. It can be shown that if the state $\psi_I(t)$ obeys the Schr\"odinger equation $i\partial_t\psi_I(t)=H_I(t)\psi_I(t)$, then the state $\psi_{II}(t)$ obeys the equation $i\partial_t\psi_{II}(t)=H_{II}(t)\psi_{II}(t)$, where $H_{II}(t)$ is given by
\begin{equation}
H_{II}(t)=U^\dagger_{II}(t)H_{I}(t)U_{II}(t) -iU^\dagger_{II}(t)\dot{U}_{II}(t)= U^\dagger_{II}(t)H_{I}(t)U_{II}(t) -H_d(t)-\nu a^\dagger a=\eta\nu(e^{i\nu a^\dagger a t}a e^{-i\nu a^\dagger a t}+{\rm H.c.})\sum_{\mu=x,y,z}f_\mu(t)S_\mu
\end{equation}
or 
\begin{equation}
H_{II}(t)=\eta\nu(ae^{-i\nu t}+a^\dagger e^{i\nu t})\sum_{\mu=x,y,z}f_\mu(t)S_\mu,
\end{equation}
where $\sum_{\mu=x,y,z}f_\mu(t)S_\mu=U^\dagger_d(t)S_zU_d(t)$. An example of the shapes of functions $f_\mu(t)$ is included in Fig.~\ref{fig:S1}.

\section*{Supplementary note 2: Second-order Hamiltonian for instantaneous pulses}
With the instantaneous-pulse approximation, we neglect $f_{x,y}(t)$ and only consider $f_z(t)$, which takes the value $1$ or $-1$ depending if the number of applied pulses is even or odd, respectively. After using the Fourier decomposition we have 
\begin{equation}
H(t)=  \eta \nu\sum_{n=1}^{\infty} f_n  \cos{(n\omega_k t)} (ae^{-i\nu t}+a^\dagger e^{i\nu t})S_z,
\end{equation}
or 
\begin{equation}
H(t)=  \frac{1}{2}\eta \nu\Big\{ f_k e^{i(k\omega_k-\nu)t} +\sum_{n\neq k}^{\infty} f_n  e^{i(n\omega_k-\nu)t}+ \sum_{n=1}^{\infty} f_ne^{-i(n\omega_k+\nu) t} \Big\} aS_z+{\rm H.c.}. 
\end{equation}
For the first term, we assume $|k\omega_k-\nu| \sim |f_k|\eta \nu/2$, while for the second and third we assume $|n\omega_k-\nu|  \gg |f_n|\eta \nu/2$ and $|n\omega_k+\nu|  \gg |f_n|\eta \nu/2$, respectively. As $\omega_k\approx\nu/k$ and $|f_n|=4/k\pi$ for instantaneous pulses, the last two conditions approximately reduce to $|n/k-1| \gg 2\eta/n\pi$ and $|n/k+1|  \gg 2\eta/n \pi$. That is, $\eta \ll \frac{n\pi}{2}|\frac{n}{k}\pm 1|$ for $n\neq k$, and $\eta \ll k\pi$. If $\eta\ll 1$, the latter is true for all $k$. For $n=1$, the former condition becomes $\eta \ll \frac{\pi}{2}|\frac{1}{k}\pm 1|\approx \frac{\pi}{2}$ which is also true if $\eta\ll 1$. For $n=k\pm 2$ (note that $f_n=0$ for $n$ even), this becomes  $\eta \ll \frac{(k\pm2)\pi}{2}|\frac{k\pm 2}{k}\pm 1|$ instead, which is true if $\eta \ll (1-2/k)\pi$. $(1-2/k)\pi$ is the smallest for $k=3$ (note that if $k=1$, $n=k-2$ makes no sense), which leads to $\eta \ll \pi/3\approx 1$. 

All in all, if $\eta\ll 1$ we can use the rotating-wave approximation and substitute the second and third terms by their respective second-order Hamiltonian. For this, we use 
\begin{equation}\label{SecondOrder}
H= \sum_j A_j^\dagger e^{i\Delta_j t} + A_j e^{-i\Delta_j t} \approx \sum_j \frac{[A_j^\dagger,A_j]}{\Delta_j}
\end{equation}
for $|| A_j || \ll |\Delta_j|$ and $ || A_j || \ll |\Delta_j- \Delta_{j'}|$  ($j\neq j'$).

\section*{ Supplementary note 3: Fourier coefficients $f_n$ and dispersive coupling $J_k$  }
In this section, we give analytical expressions for the Fourier coefficients $f_n$ for the cases of instantaneous, top-hat, and modulated pulses. For the case of instantaneous pulses, $f^{\rm \,i}_n=\frac{4}{n\pi}\sin{(n\pi/2)}$. For top-hat pulses, 
\begin{equation}
f^{\rm th}_n=\frac{4\sin{(n\pi/2)} \cos{(n\pi t_\pi/\tau)}}{n\pi(1-n^2t_\pi^2/4\tau^2)}
\end{equation}
where $t_{\pi}$ is the duration of the $\pi$-pulse, and $\tau$ is the period of the modulation function $f_z(t)$. For the modulated case, the expressions of the Fourier coefficients are
\begin{equation}
f^{\rm m}_1=1 -\frac{4d\sin(\pi k/2)}{\pi k b}\Bigg\{\frac{e^{-[(k- 1)\pi  c]^2/4}}{(k- 1)\pi}\sin{[(k-1)\pi b]} - \frac{e^{-[(k+ 1)\pi  c]^2/4}}{(k+1)\pi}\sin{[(k+1)\pi b]} \Bigg\}
\end{equation}
for $n=1$,  $f^{\rm m}_k=-4d/\pi k$ for $n=k$, and
\begin{equation}
f^{\rm m}_n=-\frac{4d\sin{(n\pi/2)}\sin(\pi k/2)}{\pi k b}\Bigg\{\frac{e^{-[(k- n)\pi  c]^2/4}}{(k- n)\pi}\sin{[(k-n)\pi b]} - \frac{e^{-[(k+ n)\pi  c]^2/4}}{(k+n)\pi}\sin{[(k+n)\pi b]} \Bigg\}.
\end{equation}
for the rest. Here, $k$ is the value of the selected harmonic, and $b, c$ and $d$ are pulse parameters. Suitable values of these can be found in Appendix~A. For the single mode case, the dispersive coupling is
\begin{equation} 
J_k= f_k^2/4 + \sum_{n\neq k}^{\infty} f_n^2/(1-n^2/k^2),
\end{equation}
and we find that its value is sufficiently accurate truncating the sum at $2k$.

\subsection{Derivations}
A periodic modulation function $f_z(t)$ can be written as a sum of infinite Fourier harmonics 
\begin{equation}
f_z(t)=\sum_{n=1}^{\infty}f_n\cos{(n\omega t)},
\end{equation}
where $\omega=2\pi/\tau$ and $f_n=\frac{2}{\tau}\int_{0}^\tau dt' f_z(t')\cos{(n\omega t')}$. To calculate the coefficients $f_n$, the change of variable $t'=x\tau/2$ is appropiate. The expression then changes to 
\begin{equation}\label{CoefficientGeneral}
f_n=\int_{0}^2 dx\ f_z(x)\cos{(n\pi x)},
\end{equation}
where $f_z(x)$ has now periodicity of $x_\tau=2$. Because $f_z(x)$ is always symmetric with respect to (w.r.t.) the $x=1$ point, i.e. $f_z(2-x)=f_z(x)$, we divide the integral in two parts
\begin{equation}
f_n=\int_{0}^1 dx\ f_z(x)\cos{(n\pi x)}+\int_{1}^2 dx\ f_z(x)\cos{(n\pi x)}.
\end{equation}
With a change of variable $x\rightarrow 2-x$ in the second integral, and using that $\cos{(\pi n (2-x))}=\cos{\pi n x}$, this results in 
\begin{equation}
f_{ n }=2\int_{0}^{1} f(x) \cos{(\pi n x)}  \ dx.
\end{equation}
Moreover, $f_z(x)$ is also antisymmetric w.r.t. the point $x=1/2$, i.e. $f_z(1-x)=-f_z(x)$. We further divide the integral, 
\begin{equation}
f_n=2\int_{0}^{1/2} dx\ f_z(x)\cos{(n\pi x)}+2\int_{1/2}^1 dx\ f_z(x)\cos{(n\pi x)},
\end{equation} 
and, with the change of variable $x\rightarrow 1-x$ in the second integral, we get
\begin{equation}
f_n=2[1-\cos{(n\pi)}]\int_{0}^{1/2} dx\ f_z(x)\cos{(n\pi x)}.
\end{equation} 
If $n$ is even, value of the integral is zero. In $n$ is odd, $f_n$ is
\begin{equation}
f_n=4\int_{0}^{1/2} dx\ f_z(x)\cos{(n\pi x)},
\end{equation} 
and its value will depend on the form of $f_z(x)$. For convenience, we make a last change of variable $x\rightarrow 1/2-x$ leading to 
\begin{equation}
f_n=4\sin{(n\pi/2)}\int_{0}^{1/2} dx\ f_z(x)\sin{(n\pi x)},
\end{equation} 
where, now, $x=0$ is the central point of the pulse. In the following we calculate $f_n$ for the cases of instantaneous, top-hat, and modulated $\pi$ pulses. \\

{ \bf Instantaneous Pulses:} For instantaneous pulses, $f_z(x)=1$, thus, $f_n=\frac{4}{n\pi}\sin{(n\pi/2)}$. \\

{ \bf Top-Hat Pulses:} For top-hat pulses, $f_z(x)=\sin{(\pi x/\tau_\pi)}$ from $x=0$ to $x=\tau_\pi/2$, while $f_z(x)=1$ for the rest. Here, $\tau_\pi=2t_\pi/\tau=2\pi/\Omega\tau$. We divide the integral accordingly, 
\begin{equation}
f_n/4\sin{(n\pi/2)}=\int_{0}^{\tau_\pi/2} dx\ \sin{(\pi x/\tau_\pi)}\sin{(n\pi x)} +\int_{\tau_\pi/2}^{1/2} dx\sin{(n\pi x)} .
\end{equation} 
For the first part, we use that $2\sin A\sin B=\cos{(A-B)}-\cos{(A+B)}$, and, then,
\begin{equation}
\int_{0}^{\tau_\pi/2} dx\ \sin{(\pi x/\tau_\pi)}\sin{(n\pi x)}=\frac{\sin{[\pi(\tau_\pi^{-1}-n)\tau_\pi/2]}}{2\pi(\tau_\pi^{-1}-n)}-\frac{\sin{[\pi(\tau_\pi^{-1}+n)\tau_\pi/2]}}{2\pi(\tau_\pi^{-1}+n)},
\end{equation}
and this first part gives
\begin{equation}
\frac{\tau_\pi}{2\pi}\Big(\frac{1}{1-n\tau_\pi} -\frac{1}{1+n\tau_\pi}\Big)\cos{(n\pi\tau_\pi/2)}.
\end{equation}
The second part gives $\frac{1}{n\pi}\cos{(n\pi\tau_\pi/2)}$. The whole integral gives
\begin{equation}
f_n=\frac{4}{n\pi}\sin{(n\pi/2)} \cos{(n\pi\tau_\pi/2)}\Big[ \frac{n\tau_\pi}{2}\frac{2n\tau_\pi}{1-n^2\tau_\pi^2} +1\Big]=\frac{4\sin{(n\pi/2)} \cos{(n\pi\tau_\pi/2)}}{n\pi(1-n^2\tau_\pi^2)},
\end{equation}
or 
\begin{equation}
f_n=\frac{4\sin{(n\pi/2)} \cos{(n\pi t_\pi/\tau)}}{n\pi(1-n^2t_\pi^2/4\tau^2)}.
\end{equation}
\\
{ \bf Modulated Pulses:} The {\it ansatz} proposed for the modulation function is 
\begin{equation}\label{Ansatz2}
f(t)=\cos[{\pi (t-t_i)/t_\pi]} + \frac{d}{\pi k b}\sin(\pi k/2)\bigg\{{\rm erf}\bigg(\frac{t-t_l}{ct_\pi}\bigg)-{\rm erf}\bigg(\frac{t-t_r}{ct_\pi}\bigg)\bigg\}\sin{[k\omega_k (t-t_m)]},
\end{equation}
where $t_\pi$ is the duration of the $\pi$ pulse, $t_i$ and $t_m=t_i+t_\pi/2$ are the initial and central points of the pulse, $t_r=t_m+bt_\pi$ and $t_l=t_m-bt_\pi$. Also, $0<b<0.5$ and $c\ll 1$. Moreover, we extend the pulse over $\tau/2$, i.e. $t_\pi=\tau/2$. The modulation function with the introduced change of variables is then
\begin{equation}\label{Ansatz3}
f_z(x)=\sin{(\pi x)} -\frac{d}{\pi k b}\sin(\pi k/2)\sin{(k \pi  x)}\, {\rm erfc}\bigg(\frac{x-b}{c}\bigg),
\end{equation}
where ${\rm erfc}(x)=1-{\rm erf}(x)$ and the Fourier coefficient $f_n$ is given by 
\begin{equation}
f^{\rm m}_n=4\sin{(n\pi/2)}\Bigg\{\int_{0}^{1/2} dx\ \sin{(\pi x)}\sin{(n\pi x)} -\frac{d}{\pi k b}\sin(\pi k/2)\int_{0}^{1/2} dx\ \sin{(k \pi  x)}\, {\rm erfc}\bigg(\frac{x-b}{c}\bigg)\sin{(n\pi x)}\Bigg\}.
\end{equation}
If $n=1$, the first part gives $1/4$. If $n\neq 1$ this is $0$. The second part can be rewritten as
\begin{equation}
-\frac{a}{2\pi k b}\sin(\pi k/2)\int_{0}^{1/2} dx\ [\cos{[(k-n) \pi  x]}-\cos{[(k+n) \pi  x]}]\, {\rm erfc}\bigg(\frac{x-b}{c}\bigg),
\end{equation}
and, for $n=k$, this simplifies to 
\begin{equation}
-\frac{d}{2\pi k b}\sin(\pi k/2)\int_{0}^{1/2} dx\ [1-\cos{(2k \pi  x)}]\, {\rm erfc}\bigg(\frac{x-b}{c}\bigg).
\end{equation}
If $1/2-b\gg c$, then $ {\rm erfc}[(1/2-b)/c]\approx0$ and the upper limit of the integral can taken to be $\infty$. With a last change of variables $x\rightarrow x+b$, the second part is
\begin{equation}
-\frac{d}{2\pi k b}\sin(\pi k/2)\int_{-b}^{\infty} dx\ \{\cos{[\pi (k-n) (x+b)]}-\cos{[\pi(k+n) (x+b)]}\}\, {\rm erfc}(x/c).
\end{equation}
and, for $n=k$,
\begin{equation}
-\frac{d}{2\pi k b}\sin(\pi k/2)\int_{-b}^{\infty} dx\ \{1-\cos{[2k \pi  (x+b)]}\}\, {\rm erfc}(x/c).
\end{equation}
In the appropriate regime, these integrals result in~\cite{Ng68}
\begin{equation}
-\frac{d\sin(\pi k/2)}{\pi k b}\Bigg\{\frac{e^{-[(k- n)\pi  c]^2/4}}{(k- n)\pi}\sin{[(k-n)\pi b]} - \frac{e^{-[(k+ n)\pi  c]^2/4}}{(k+n)\pi}\sin{[(k+n)\pi b]} \Bigg\}.
\end{equation}
and
\begin{equation}
f^{\rm m}_k\approx-\frac{4d}{\pi k },
\end{equation}
respectively.


\section*{ Supplementary note 4: Residual spin-phonon entanglement and micromotion}

We assume a generic multiqubit-boson Hamiltonian of the form 
\begin{equation}
H(t)=\eta\nu[ a S^\dagger(t)+a^\dagger S(t) ]
\end{equation}
In this case, the first-order unitary operator is $U(t)=\exp{[-i \int_0^t H(t') dt']}=\exp{\{ \eta [a\, \tilde{S}^\dagger(t)+a^\dagger \,\tilde{S}(t)]\}}$. We assume the ideal pure state $\rho_i$ is given after applying the second-order evolution operator $U_g=e^{i\theta(t_g)S_z^2}$. If we assume $[S_z^2, H(t)]=0$, the reduced final state is approximately given (considering only the first and second-order evolution operators) $\rho={\rm Tr}_a\{U(t)U_g \rho_0\otimes\rho_{a}U_g^\dagger U^\dagger(t)\}={\rm Tr}_a\{U(t)\, \rho_i\otimes\rho_{a}U^\dagger(t)\}$, where $t$ is now the final time $t_g$. If $\eta\ll 1$, we can expand $U(t)$ in $\eta$, and we get 
\begin{eqnarray}
\rho_E= \rho_i+ \eta^2 (\langle n\rangle+1) \Big[\tilde{S}(t) \rho_i \tilde{S}^\dagger(t) -\frac{1}{2}\{\tilde{S}^\dagger(t) \tilde{S}(t)\rho_i + \rho_i \tilde{S}^\dagger(t) \tilde{S}(t)\}\Big] + \eta^2 \langle n\rangle \Big[\tilde{S}^\dagger(t) \rho_i \tilde{S}(t) -\frac{1}{2}\{\tilde{S}(t) \tilde{S}^\dagger(t)\rho_i + \rho_i \tilde{S}(t) \tilde{S}^\dagger (t)\}\Big]
\end{eqnarray}
where we use ${\rm Tr}_a(a^n \rho_a)={\rm Tr}_a((a^\dagger)^n \rho_a)=0$, and where $\tilde{S}(t)=i\nu \int_0^t S(t') dt'$. 
Expanding the fidelity, defined here as $F=|{\rm Tr}(\rho_i\rho_E)|/\sqrt{\rho_E^2}$, in powers of $\eta$, this becomes
\begin{eqnarray}
F\approx 1-\frac{1}{2}|{\rm Tr}[(\rho_E-\rho_i)^2]|
\end{eqnarray}
In our case, $\tilde{S}(t)=i\nu \int_0^t e^{i\nu t'}f_z(t') dt' S_z=-\alpha(t)S_z/\eta $, thus, 
\begin{eqnarray}
\rho_E-\rho_i=   (2\bar{n}+1)|\alpha(t_{\rm g})|^2 \Big[S_z \rho_i S_z -\frac{1}{2}(S^2_z \rho_i +  \rho_i S^2_z )  \Big],
 \end{eqnarray}
and 
\begin{eqnarray}
{\rm Tr}[(\rho_E-\rho_i)^2]=   (2\bar{n}+1)^2|\alpha(t_{\rm g})|^4 \Big( \frac{3}{2}\langle S_z^2 \rangle^2  + \frac{1}{2}\langle S_z^4 \rangle- 2 \langle S_z^3\rangle \langle S_z\rangle    \Big),
 \end{eqnarray}
 where $\langle \cdot \rangle={\rm Tr}(\cdot \rho_i)$. For final state $|\Phi^+\rangle=\frac{1}{\sqrt{2}}(|++\rangle+i|--\rangle)$, $\langle S_z\rangle=0$, $\langle S_z^2\rangle=2$, $\langle S_z^3\rangle=0$, and  $\langle S_z^4\rangle=8$, leading to 
\begin{eqnarray}\label{ContributionInfi}
F\approx 1-5(2\bar{n}+1)^2|\alpha(t_{\rm g})|^4 
\end{eqnarray}

This result can be easily extended case to the case where the ions couple diferently to the mode, i.e. $S_z\rightarrow S_z=(\eta_1\sigma_1^z+\eta_2\sigma_2^z)/\eta$, obtaining 
 \begin{eqnarray}
F\approx 1-5(2\bar{n}+1)^2|\alpha_1(t_{\rm g})|^2 |\alpha_2(t_{\rm g})|^2,
\end{eqnarray}
where $\alpha_j(t_{\rm g})=-i\eta_j \nu\int_0^{t_{\rm g}}dt'f(t')e^{i\nu t'}$. Furthermore, for systems coupled to M modes, i.e.
\begin{equation}
H^I(t)=\sum_m\eta_m\nu_m[ a_m S_m^\dagger(t)+a_m^\dagger S_m(t) ],
\end{equation}
this fidelity bound is
 \begin{eqnarray}
F\approx 1-\sum_{m=1}^{M}5(2\bar{n}_m+1)^2|\alpha_{1m}(t_{\rm g})|^2 |\alpha_{2m}(t_{\rm g})|^2,
\end{eqnarray}
where $\alpha_{jm}(t_{\rm g})=-i\eta_{jm} \nu_m\int_0^{t_{\rm g}}dt'f(t')e^{i\nu_m t'}$, $\nu_m$ being the $m$-th mode frequency, and $\eta_{jm}$ the coupling between qubit $j$ and mode $m$.

\subsection{Intrinsic gate error: Single mode scenario}
Numerically, $\alpha(t_{})$ is can be exactly calculated solving the integral $\alpha(t_{\rm g})=-i\eta \nu\int_0^{t_{\rm g}}dt'f(t')e^{i\nu t'}$. Analytically, we estimate its value by assuming that $\alpha(t_{\rm g})=-i\eta \nu\int_0^{t_{\rm g}}dt'f(t')e^{i\nu t'}\approx-i\eta \nu\int_0^{t_{\rm g}}dt'e^{i\nu t'} \lesssim 2\eta$. This is justified as long as $|\nu -n\omega_k |t_{\rm g}\gg 1$ for all frequency components of $f(t)$. The latter condition is not true for $n=k$, however, $\int_0^{t_{\rm g}}dt'e^{-ik\omega_k t'}e^{i\nu t'}=0$ by design. Thus, an approximate upper bound for the fidelity is
\begin{eqnarray}
F\gtrsim 1-5(2\bar{n}+1)^22^4\eta^4.
\end{eqnarray}

\subsection{Trapped ions: Micromotion}
Trapped ions are often trapped in non-harmonic potentials. In the case of linear Paul traps, the dynamics of longitudinal motional modes is described by mode functions~\cite{Bermudez17}
\begin{equation}
u_{m}(t)=\frac{e^{i\nu_{m} t}}{\tilde{\xi}_{z}}\Big(1+\sum_{l \geq 1} \frac{(-1)^l2q_z^l}{4^l((l-1)!)^2}\cos{(l\Omega_{\rm rf}t)}\Big),
\end{equation}
where $\nu_{m}$ is the $m$-th mode's (secular) frequency, and $\Omega_{\rm rf}$ is the trap rf frequency associated to the so-called micromotion. Also, $\tilde{\xi}_z=1+\sum_{l\geq1}(-1)^l \frac{2q_z^l}{4^l((l-1)!)^2}$. Instead of $\alpha(t)$, the phase space trajectory of the center-of-mass mode that accounts for micromotion will be
\begin{equation}
\alpha(t)=-i\eta \nu\int_0^{t}dt'f(t')\frac{e^{i\nu t'}}{\tilde{\xi}_{z}}\Big(1+\sum_{l \geq 1} \frac{(-1)^l2q_z^l}{4^l((l-1)!)^2}\cos{(l\Omega_{\rm rf}t')}\Big)
\end{equation}
Typically, $\Omega_{\rm rf}/\nu >10$ and $q_z\sim 10^{-4}$~\cite{Bermudez17}, thus, we only keep the first term in the summation 
\begin{equation}
\alpha(t)=-i\eta \nu\int_0^{t}dt'f(t')\frac{e^{i\nu t'}}{\tilde{\xi}_{z}}\Big(1- \frac{q_z}{2}\cos{(\Omega_{\rm rf}t')}\Big).
\end{equation}
We consider a worst-case scenario in which one of the harmonics of $f(t)$ equals the frequency $\Omega_{\rm rf}-\nu$. If $k\omega_k \approx \nu$ and $n'\omega_k\approx \Omega_{\rm rf}-\nu$, the relation $\Omega_{\rm rf}/\nu \sim 10$ leads to $n'/k\sim 9$. That is, the harmonic $n'$ that enters into resonance with frequency $\Omega_{\rm rf}-\nu$ must be approximately an order of magnitude higher than the harmonic $k$. If $n'\omega_k=\Omega_{\rm rf}-\nu$, at the end of the gate, this unwanted resonance will produce a displacement of the order of 
\begin{equation}
|\alpha_{\rm mi}(t_{\rm g})| \lesssim \frac{\eta \nu q_z f_{n'}}{8}t_{\rm g}.
\end{equation}
Assuming $t_{\rm g}= 2\pi/\xi_k \sim \pi/(\eta \nu f_k)$ and $f_n\sim4/(\pi n)$, this becomes $|\alpha_{\rm mi}(t_{\rm g})|\sim \frac{\pi q_z k}{8n'}\sim \frac{\pi q_z}{8\times9}\sim 10^{-5}$. Using Eq.~(\ref{ContributionInfi}), one concludes that the contribution to the infidelity is negligible. For transversal motional modes, the situation could be different, as for these modes $q_{x,y}\sim 0.1$. However, in the implementation we considered, these modes do not play a role as they do not couple to the internal degrees of freedom.


\section*{ Supplementary note 5: Second-order Hamiltonian for non-instantaneous pulses}
\begin{figure}
\centering
\includegraphics[width=0.7\linewidth]{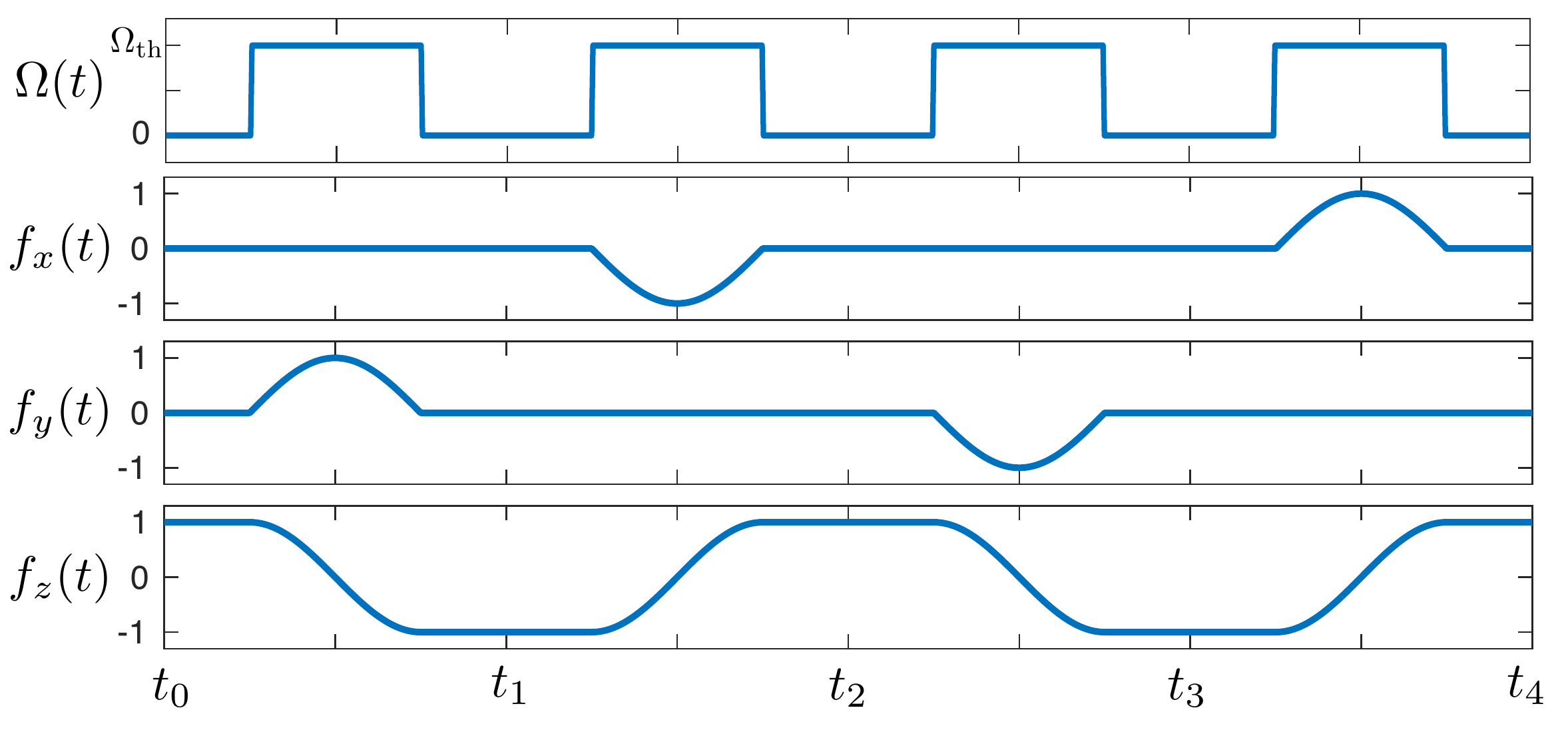}
\caption{ Form of the modulation functions $f_{x,y,z}(t)$ for a single XY4 block with top hat pulses and with $t_\pi=\tau/4=\pi/\Omega_{\rm th}$. In the XY4 block, the first and third pulse rotate the qubit along the x axis, that is, $\Omega_x(t)=\Omega(t)$, $\Omega_y(t)=0$. For the second and fourth pulses, the same is true for the y axis.}\label{fig:S1}
\end{figure}

Here we derive the second-order Hamiltonian associated to a sequence concatenating $N_{XY4}$ XY4 blocks, where each block is composed by four pulses with alternating phases, i.e. XYXY. The evolution of each block can be divided in four parts. During the first part, $U_d(t)=\exp{[-i/2 \int_{t_0}^t\Omega(t) S_x]}$, and, thus, 
\begin{equation}\label{XY4Hamil1}
H(t)=\eta \nu (ae^{-i\nu t}+a^\dagger e^{i\nu t})[f(t,t_0)S_z+f_\perp(t,t_0)S_y],
\end{equation}
where $f(t,t_0)=\cos{[\int_{t_0}^t \Omega(t',t_0)dt']}$ and $f_\perp(t,t_0)=\sin{[\int_{t_0}^t \Omega(t',t_0)dt']}$. During the second part we have that $U_d(t)=\exp{[-i/2 \int_{t_1}^t\Omega(t,t_1) S_y]}(-\sigma_1^x\sigma_2^x)$ and 
\begin{equation}\label{XY4Hamil2}
H(t)=\eta \nu (ae^{-i\nu t}+a^\dagger e^{i\nu t})[-f(t,t_1)S_z-f_\perp(t,t_1)S_x].
\end{equation}
For the third and fourth parts, we have that $U_d(t)=\exp{[-i/2 \int_{t_2}^t\Omega(t,t_2) S_x]}(-\sigma_1^z\sigma_2^z)$ and $U_d(t)=\exp{[-i/2 \int_{t_3}^t\Omega(t,t_3) S_y]}(-\sigma_1^y\sigma_2^y)$, respectively. The corresponding Hamiltonians are
\begin{equation}\label{XY4Hamil3}
H(t)=\eta \nu (ae^{-i\nu t}+a^\dagger e^{i\nu t})[-f(t,t_2)S_z-f_\perp(t,t_2)S_y] \, \, \, \,{\rm and} \, \, \, \, H(t)=\eta \nu (ae^{-i\nu t}+a^\dagger e^{i\nu t})[-f(t,t_3)S_z+f_\perp(t,t_3)S_x],
\end{equation}
respectively. When concatenating $N_{XY4}$ blocks where $t_0=0, t_1=\tau/2,  t_2=\tau, t_3=3\tau/2$ and $t_4=2\tau$, one can use the following Hamiltonian to describe the dynamics
\begin{equation}\label{XY4Hamil}
H(t)=\eta \nu (ae^{-i\nu t}+a^\dagger e^{i\nu t})[f_x(t)S_x+f_y(t)S_y+f_z(t)S_z],
\end{equation}
where
\begin{eqnarray}\label{modulx}
f_{x}(t)&=&\sum_{n=1}^{N}-f_\perp(t,t_1+2n\tau)\Big\{\Theta[t-t_1-2n\tau]-\Theta[t-t_2-2n\tau]\Big\} 
+f_\perp(t,t_3+2n\tau)\Big\{\Theta[t-t_3-2n\tau]-\Theta[t-t_4-2n\tau]\Big\}, \\
f_{y}(t)&=&\sum_{n=1}^{N}f_\perp(t,2n\tau)\Big\{\Theta[t-2n\tau]-\Theta[t-t_1-2n\tau]\Big\} -f_\perp(t,t_2+2n\tau)\Big\{\Theta[t-t_2-2n\tau]-\Theta[t-t_3-2n\tau]\Big\}\nonumber ,
\end{eqnarray}
and
\begin{eqnarray}\label{modulz}
f_{z}(t)&=&\sum_{n=1}^{2N}f(t,n\tau)\Big\{\Theta[t-n\tau]-\Theta[t-t_1-n\tau]\Big\} -f(t,t_1+n\tau)\Big\{\Theta[t-t_1-n\tau]-\Theta[t-t_2-n\tau]\Big\} ,\nonumber
\end{eqnarray}
where $\Theta(t-t_i)$ is the Heaviside step function centred in $t_i$. The form of the modulation functions defined in Eqs.~(\ref{modulx})-(\ref{modulz}) is shown in Fig.~\ref{fig:S1} for the case of top-hat $\pi$ pulses with length $t_\pi=\tau/4$.

For sequences built by concatenating XY4 blocks, the periods of $f_{z}(t)$ and $f_{x,y}(t)$ differ by a factor of two. After applying the Fourier decomposition to Eqs.~(\ref{modulx})-(\ref{modulz}), these become
\begin{equation}\label{Fmodulx}
f_{x}(t)=\sum_{n=1}^{\infty}a^x_n\cos{(n\omega t/2)}+b^x_n\sin{(n\omega t/2)}, \, \, \, \, \, \, \, \, \, \, \, \, f_{y}(t)=\sum_{n=1}^{\infty}a^y_n\cos{(n\omega t/2)}+b^y_n\sin{(n\omega t/2)},
\end{equation}
and
\begin{equation}\label{Fmodulz}
f_{z}(t)=\sum_{n=1}^{\infty}f_n\cos{(n\omega t)},
\end{equation}
where now $n$ stands for the number of the Fourier harmonic, $\omega=2\pi/\tau$, and the Fourier coefficients are given by 
\begin{equation}\label{acoeff}
a^{x,y}_n=\frac{1}{\tau}\int_0^{2\tau}f_{x,y}(t)\cos{(n\omega t/2)}, \, \, \, \, \, \, \, \, \, \, \, \,  b^{x,y}_n=\frac{1}{\tau}\int_0^{2\tau}f_{x,y}(t)\sin{(n\omega t/2)},
\end{equation}
and
\begin{equation}\label{zcoeff}
f_n=\frac{2}{\tau}\int_0^{\tau}f_{z}(t)\cos{(n\omega t)}.
\end{equation}
Here, the sine components of $f_z(t)$ are zero because the integral $\int_0^{\tau}f_{z}(t)\sin{(n\omega t)}$ is zero for all $n\in\mathbb{N}$. This can be proven by dividing the integral in two parts ($\{0,t_1\}$ and $\{t_1,t_2\}$) and using the symmetry property $f_z(t_1+t)=f_z(t_1-t)$. In a similar fashion, the symmetry properties $f_y(t+t_2)=-f_y(t)$, $f_x(t+t_2)=-f_x(t)$, $f_y(t_1/2-t)=f_y(t_1/2+t)$, and $f_x(3t_1/2-t)=f_x(3t_1/2+t)$ along with trigonometric identities $\sin{(a\pm b)}=\sin{a}\cos{b}\pm\cos{a}\sin{b}$ and $\cos{(a\pm b)}=\cos{a}\cos{b}\mp\sin{a}\sin{b}$, can be used to prove that
\begin{eqnarray}
a^{x}_n&=&-\cos{(3\pi n/4)}e_n, \ \ b^{x}_n=-\sin{(3\pi n/4)}e_n, \\
a^{y}_n&=&\cos{(\pi n/4)}e_n, \  \ \ \ b^{y}_n=\sin{(\pi n/4)}e_n,
\end{eqnarray}
where $e_{n}=[\cos{(n \pi )}-1]\frac{2}{\tau}\int_0^{t_\pi/2}f_\perp(t+t_1/2,0)\cos{(n\omega t /2)}dt$. Note that $e_n$ is zero if $n$ is even. 

In the following, we will calculate the second-order Hamiltonian associated to Eq.~(2) of the main text after a XY4 pulse sequence. After the Fourier expansion, this looks like
\begin{equation}
H=\frac{1}{2}\eta\nu a^\dagger e^{i\nu t}\sum_{n=1}^{\infty} \Big\{f_n(e^{in\omega t} +e^{-in\omega t})S_z +(\bar{c}_{n}^xS_x+\bar{c}_{n}^yS_y)e^{in\omega t/2} +({c}_{n}^xS_x+{c}_{n}^yS_y)e^{-in\omega t/2}\Big\} +{\rm H.c.},
\end{equation}
where $c_n^{x,y}=a_n^{x,y}+ib_n^{x,y}$, and $\bar{c}_n^{x,y}$ is the complex conjugate. As $e_n$ is zero for even $n$, $n\omega$ and $n\omega/2$ never coincide, and, if $\eta\nu \ll \omega$, the second order Hamiltonian is (see Eq.~(\ref{SecondOrder}))
\begin{equation}\label{soHamil}
H_{\rm XYXY}^{(2)}=\frac{1}{2}\eta^2\nu \Big\{ -J_k^\parallel S_z^2 - J_k^\perp (S_x^2 + S_y^2)  + [B_k' (a^\dagger a+1) +B_k'']S_z  + J_k^{xy}(\sigma_1^x\sigma_2^y+\sigma_1^y\sigma_2^x)\Big\} ,
\end{equation}
where 
\begin{eqnarray}\label{Couplings}
J_k^\parallel&=& f_k^2/2 +\sum_{n\neq k}^{\infty}\frac{ f_n^2}{1-n^2\omega^2/\nu^2},\label{Couplings1} \\
J_k^\perp&=& \frac{1}{2}\sum_{n=1}^{\infty}\frac{ e_n^2}{1-n^2\omega^2/4\nu^2},\label{Couplings2}\\
B_k'&=&2\sum_{n=1}^{\infty}\frac{n\omega}{\nu}\frac{ e_n^2}{1-n^2\omega^2/4\nu^2}\sin{(n\pi /2)}, \\
B_k''&=&\sum_{n=1}^{\infty}\frac{ e_n^2}{1-n^2\omega^2/4\nu^2}\sin{(n\pi /2)}, \\
J_{k}^{xy}&=&\sum_{n=1}^{\infty}\frac{ e_n^2}{1-n^2\omega^2/4\nu^2}\cos{(n\pi /2)}. 
\end{eqnarray}

As we indicated above, $e_n$ is zero for even $n$. Thus, $J_k^{xy}=0$ for XY4 sequences. The rest of the terms will have a non-zero value and ought to be removed by applying refocusing techniques. The second-order Hamiltonian for the inverse XY4 sequence, that is, YXYX, is equivalent except for the term $\frac{1}{2}\eta^2\nu[B_k' (a^\dagger a+1) +B_k'']S_z$, which has the opposite sign. Concatenating the XYXY and the YXYX blocks in what is called the XY8 sequence, we will achieve a partial refocusing of this term. Refocusing the $-\frac{1}{2}\eta^2\nu J_k^\perp(S_x^2+S_y^2)$ term is possible by driving the two qubits with opposite phases in a second XY8 block. The effective Hamiltonian will be equivalent except that, now, the operators $S_{x,y}$ will become $S_{x,y}^{(-)}=\sigma_1^{x,y}-\sigma_2^{x,y}$, changing the sign of terms $\sigma_1^{u}\sigma_2^{\mu}$ where $u,\mu\in\{x,y\}$.

\section*{  Supplementary note 6: Extension to two-qubit and multi-qubit gates in the absence of individual addressing} 

In the absence of individual addressing, one is not able to change the relative phase between the qubit drivings. As a consequence, the term $J_k^\perp(S_x^2+S_y^2)$ in Eq.~(10) can not be refocused and the applied gate operation at time $t_{\rm g}$ is, instead of $U(t_{\rm g})\approx \exp{[i \theta(t_{\rm g})S_z^2]}$, 
\begin{equation}\label{fsimuni}
V(t_{\rm g})\approx  e^{i \theta(t_{\rm g})S_z^2}e^{i\chi(t_{\rm g}) (S_x^2+S_y^2)},
\end{equation}
where $\chi(t_{\rm g})=\frac{1}{2}\eta^2\nu J_k^\perp t_{\rm g}$ or
\begin{equation}
\chi(t_{\rm g})=\frac{\pi \eta J_k^\perp}{2\Big\{\sqrt{f_k^2+4\eta^2J_k^2}+2\eta J_k\Big\}}.
\end{equation}
Note that $S_z^2$ and $(S_x^2+S_y^2)$ commute, thus, the evolution operator $V(t_{\rm g})$ can be written as the product of $U(t_{\rm g})$ and $e^{i\chi (S_x^2+S_y^2)}$. Also, notice that the value of $\chi$ can be numerically determined using Eqs.~(\ref{Couplings1}) and (\ref{Couplings2}). Typically, we expect $\chi<\pi/8$.

The $U(t_{\rm g})$ relates, up to a global phase and qubit rotation, with the so-called CPHASE gate, i.e. $U_{\rm CPHASE}=\exp{(-i\frac{\pi}{4})}\exp{(-i\frac{\pi}{4}S_z)}U(t_{\rm g})$. The unitary operator in Eq.~(\ref{fsimuni}) holds a similar relation with the so-called ``fSim gate" fSim$(\varphi,\phi)$ for the case $\phi=\pi$~\cite{Foxen20} and where $\theta$ is proportional to $\chi$. Note that two fSim$(\theta,\pi/2)$ gates combined with a local operation in one of the qubits will lead to $U_{\rm CPHASE}$ gate (up to a global phase and qubit rotation), while for generating a fSim$(\theta,\pi/2)$ gate, we have to change the condition $\theta(t_{\rm g})=\pi/8$ for $\theta(t_{\rm g})=\pi/16$ instead.

For the multiqubit case, we note that the only step of our protocol that requires $S_{x,y,z}$ to be composed by two-qubit operators, i.e. $S_{\alpha}=\sigma_1^\alpha+\sigma_2^{\alpha}$, is the refocusing of the $J_k^\perp(S_x^2+S_y^2)$ in Eq.~(10). In the absence of this step, the extension to the multiqubit case is trivial just by replacing all $S_{\alpha}=\sigma_1^\alpha+\sigma_2^{\alpha}$ operators by $S_{\alpha}=\sum_{j=1}^{N_q}\sigma_j^\alpha$, where $N_q$ is the number of qubits. In this case, and unlike for two qubits, a trivial generalisation to inhomogeneous qubit-boson couplings is not possible. Still, $S_z^2$ and $(S_x^2+S_y^2)$ commute, thus, the evolution operator at the end of the gate will be given by Eq.~(\ref{fsimuni}). In order to relate Eq.~(\ref{fsimuni}) to a known multiqubit operation, we use that $[S^2,S_z]=0$, where $S^2=S_x^2+S_y^2+S_z^2$ to rewrite it as
\begin{equation}
V(t_{\rm g})\approx  e^{i [\theta(t_{\rm g})- \chi(t_{\rm g})]S_z^2}e^{i\chi(t_{\rm g}) S^2}.
\end{equation}
From Ref.~\cite{Sorensen00_2}, we know that the application of $U(t_{\rm g})=e^{i\frac{\pi}{8} S_z^2}$ to state $|-\rangle^{\otimes N_q}$ generates the GHZ state $|\Psi\rangle=\frac{1}{\sqrt{2}}[|-\rangle^{\otimes N_q}-i|+\rangle^{\otimes N_q}]$ for $N_q$ even.
The action of $V(t_{\rm g})$ in state $|-\rangle^{\otimes N_q}$ will have a similar effect. Note that $e^{i\chi(t_{\rm g}) S^2}$ acts trivially in state $|-\rangle^{\otimes N_q}$, adding a global phase factor. As already discussed, the first part $e^{i [\theta(t_{\rm g})- \chi(t_{\rm g})]S_z^2}$ will transform the state $|-\rangle^{\otimes N_q}$ into $|\Psi\rangle$ for $N_q$ even, if  $\theta(t_{\rm g})- \chi(t_{\rm g})=\pi/8$. This could be achieved by modifying the condition $\theta(t_{\rm g})=\pi/8$ targeted in the main text by $\theta(t_{\rm g})=\pi/8+ \chi(t_{\rm g})$. As we do not provide an analytic expression for $J_k^\perp$ (which determines $\chi$), a value for the detuning $\xi_k$ that that fulfils condition $\theta(t_{\rm g})=\pi/8+ \chi(t_{\rm g})$ will have to be found using numerical integration.

\section*{ Supplementary note 7: Details of numerical simulations}

\subsection{Two-mode Hamiltonian}

The Hamiltonian describing two trapped ions under a static magnetic field gradient $\partial B/\partial z=g_B$, in an interaction picture w.r.t. free energy term of the qubits  $\frac{\omega_1}{2}\sigma_1^z+\frac{\omega_2}{2}\sigma_2^z$ is~\cite{Arrazola18_2}
\begin{equation}\label{Hzero2ions}
H_{0}+H_{\rm 2M}= \nu a^\dagger a+\sqrt{3}\nu b^\dagger b +\eta\nu(a+a^\dagger)S_z^{(+)} -3^{-1/4}\eta\nu(b+b^\dagger)S_z^{(-)},
\end{equation}
where $a^\dagger$($a$) and $b^\dagger$ ($b$) are the creation (annihilation) operators associated to the longitudinal center-of-mass and breathing modes of the two-ion crystal. To simulate the performance of the gate under the influence of two modes, we evolve the state according to Hamiltonian $H_f=H_d^{(\pm)}+H_{0}+H_{\rm 2M}$, starting from motional thermal states with $\bar{n}=1$, and truncating the Hilbert space of the first ($a$) and second ($b$) mode to $16$ and $12$ states, respectively.

\subsection{Effective Hamiltonian of the driven system}

Here we justify the use of the term $H_{\rm 2M}^{\rm eff}$ as a right quantity to account for the second-order effect of the breathing mode, without explicitly taking it into account in our simulations. For that, we first calculate the effective Hamiltonian of the driven system for the two-mode case.

After moving into an interaction picture w.r.t. $H_d+ \nu a^\dagger a + \sqrt{3}\nu b^\dagger b$, neglecting the terms going with $f_{x,y}(t)$ and expanding $f_z(t)$ as a Fourier series, the Hamiltonian of the driven, two-mode system is
\begin{equation}\label{Hdriven2ions}
H'(t)=\sum_{n=1}^{\infty} \frac{1}{2}f_n\eta \nu S_z^{(+)} ae^{-i\nu t}(e^{in\omega_k t}+e^{-in\omega_k t}) - \frac{3^{-1/4}}{2}f_n\eta\nu S_z^{(-)}be^{-i\sqrt{3}\nu t}(e^{in\omega_k t}+e^{-in\omega_k t})  +{\rm H.c.}.
\end{equation}
In Hamiltonian~(\ref{Hdriven2ions}), the only resonant term is the one going with $e^{-i(\nu-k\omega_k)}$ and its complex conjugate, i.e.  $\frac{1}{2}f_k\eta \nu S_z a e^{-i(\nu-k\omega_k) t}+ {\rm H.c.}$. Using Eq.~(\ref{SecondOrder}), the influence of the rest of the terms is well approximated by the second order Hamiltonian
\begin{equation}\label{Hsecond2ions}
H_{\rm eff}= -\frac{1}{2}\eta^2\nu J^{a}_k(S_z^{(+)})^2 -\frac{1}{6}\eta^2\nu J^{b}_k(S_z^{(-)})^2,
\end{equation}
where $J^{a}_k=f_k^2/4+\sum_{n\neq k}^{\infty} f_n^2/(1-n^2/k^2)$ and $J^{b}_k=\sum_{n=1}^{\infty} f_n^2/(1-n^2/3k^2)$. Because $(S_z^{(-)})^2=4\mathbb{I}-S_z^2$, Hamiltonian~(\ref{Hsecond2ions}) is, up to a global phase,
\begin{equation}\label{Hsecond2ions_2}
H_{\rm eff}= -\frac{1}{2}\eta^2\nu (J^{a}_k- J^{b}_k/3)S_z^2,
\end{equation}
The effective Hamiltonian describing the driven system is then
\begin{equation}\label{Hdrivenplussecond}
H'(t)\approx \frac{1}{2}f_k\eta \nu S_z a( e^{-i(\nu-k\omega_k) t}+ {\rm H.c.})-\frac{1}{2}\eta^2\nu (J^{a}_k- J^{b}_k/3)S_z^2,
\end{equation}
which is equivalent to Hamiltonian~(4) in the main text but with $J_k\rightarrow J_k- J^{b}_k/3$. In our numerical simulations, we do not want to include the dynamics of the breathing bosonic mode. Yet, we want to capture its second-order effect correctly. This is achieved by the following Hamiltonian 
\begin{equation}\label{Hions}
H=H_{d} + \nu a^\dagger a + \eta \nu (a+a^\dagger)S_z +\frac{1}{3}\nu \eta^2 r S_z^2,
\end{equation}
where $r=J_k^b/\sum_{n=1}^{\infty}f_n^2$. Following the procedure described from Eq.~(\ref{Hzero2ions}) to Eq.~(\ref{Hdrivenplussecond}), but now with $H$, one realises that the effective Hamiltonian corresponding Eq.~(\ref{Hions}) is equivalent to $H'(t)$.

\subsection{Crosstalk}

In the two-ion system, the frequency of qubit $j$ is related with the intensity of the magnetic field in the ion's equilibrium position $z_j^0$, and both frequencies differ by $\Delta\omega=\omega_2-\omega_1=\gamma_eg_Bd$. Here, the distance between the ions is $d=(e^2/2\pi\varepsilon_0 M\nu^2)^{1/3}$ where $e$ is the electric charge of the electron and $\varepsilon_0$ is the vacuum permittivity. For two $^{171}$Yb$^{+}$ ions each with mass $M=171$~amu and trapped with longitudinal frequency $\nu=(2\pi)\times 220$ kHz, this value gives $\Delta\omega=2.54$ and $20.34$ MHz for $g_B=19.57$ and $ 153.2\,\rm T/m$, respectively. 

Assuming we address only two levels of each ion, the Hamiltonian of the driven qubits is  
\begin{equation}\label{TotalDriving}
H_{dq}=\frac{\omega_1}{2}\sigma_1^z+\frac{\omega_2}{2}\sigma_2^z+ \Omega(t)(\sigma_1^x+\sigma_2^x)\cos{(\omega_1 t-\phi_1)}+ \Omega(t)(\sigma_1^x+\sigma_2^x)\cos{(\omega_2 t-\phi_2)},
\end{equation}
where $\Omega_1=\Omega_2$ for simplicity, and $\phi_j$ is the phase of driving $j$. In an interaction picture w.r.t.  $\omega_1/2\sigma_1^z+\omega_2/2\sigma_2^z$, and after neglecting fast oscillating terms, the Hamiltonian reads

\begin{equation}\label{TotalDrivingRWA}
{H}_{dq}^I\approx \frac{\Omega(t)}{2}(\sigma_1^+e^{i\phi_1} + \sigma_2^+e^{i\phi_1}e^{-i\Delta\omega t} +{\rm H.c.} )  +\frac{\Omega(t)}{2}(\sigma_2^+e^{i\phi_2}  +\sigma_1^+e^{i\phi_2}e^{i\Delta\omega t} +{\rm H.c.} ) 
\end{equation}
Notice that, if $\Omega(t)\ll\Delta\omega$ , the terms oscillating with $\Delta\omega$ can also be neglected. This means that, in this regime, the ions can be then selectively addressed with global fields. For the case $\phi_1=\phi_2=\phi$, Hamiltonian~(\ref{TotalDrivingRWA}) becomes
\begin{equation}\label{HdrivingwCross_1}
{H}_{dq}^I\approx\frac{\Omega_x(t)}{2}S^{(+)}_x +\frac{\Omega_x(t)}{2}(\sigma_2^-e^{-i\Delta\omega t} +\sigma_1^-e^{i\Delta\omega t} +{\rm H.c.} ) 
\end{equation}
or 
\begin{equation}\label{HdrivingwCross_2}
{H}_{dq}^I\approx \frac{\Omega_y(t)}{2}S_y^{(+)}  + \frac{\Omega_y(t)}{2} (i\sigma_2^-e^{-i\Delta\omega t} +i\sigma_1^-e^{i\Delta\omega t}+{\rm H.c.} ),
\end{equation}
depending $\phi=0$ or $\pi/2$. In the main text, we describe both cases using the more general Hamiltonian 
\begin{equation}\label{HdrivingwCrossplus}
H_{d}^{(+)}+H_c^{(+)}= \frac{\Omega_x(t)}{2}S^{(+)}_x + \frac{\Omega_y(t)}{2}S_y^{(+)}  + \frac{\Omega_x(t)}{2}(\sigma_2^-e^{-i\Delta\omega t} +\sigma_1^-e^{i\Delta\omega t} +{\rm H.c.} ) + \frac{\Omega_y(t)}{2} (i\sigma_2^-e^{-i\Delta\omega t} +i\sigma_1^-e^{i\Delta\omega t}+{\rm H.c.} ),
\end{equation}
where $\Omega_x(t)=0$ ($\Omega_y(t)=0$) when $\phi=\pi/2$ ($\phi=0$). For the case where the qubits are driven with opposite phase, i.e. $\phi_2=\phi_1+\pi=\phi+\pi$, Hamiltonian~(\ref{HdrivingwCrossplus}) is
\begin{equation}\label{HdrivingwCrossminus}
H_{d}^{(-)}+H_c^{(-)}= \frac{\Omega_x(t)}{2}S^{(-)}_x + \frac{\Omega_y(t)}{2}S_y^{(-)}  + \frac{\Omega_x(t)}{2}(\sigma_2^-e^{-i\Delta\omega t} -\sigma_1^-e^{i\Delta\omega t} +{\rm H.c.} ) + \frac{\Omega_y(t)}{2} (i\sigma_2^-e^{-i\Delta\omega t} -i\sigma_1^-e^{i\Delta\omega t}+{\rm H.c.} )
\end{equation}
instead. The Hamiltonian used for the simulations with crosstalk are then $H_s+H^{(+)}_d+H^{(+)}_c$  and $H_s+H^{(-)}_d+H^{(-)}_c$, for every first and second half of a TQXY16 block, respectively. Doing so, we assume that the phase of the drivings can be changed instantaneously.

The results of the simulations show that, for the Rabi frequencies we use, crosstalk terms have a non-negligible effect. To reduce their impact we combine each pulse with a sin$^2$-shaped ramp at the beginning and end of each pulse. Each pulse is then constructed following 
\begin{equation}
\Omega(t)=
        \begin{cases}
         -\frac{\partial f_z(0)}{\partial t}\times[1-f^2_z(0)]^{-1/2} \times \sin^2({\pi t/2t_{\rm ramp}}) \,\,\, {\rm if} \,\,\, 0<t\leq t_{\rm ramp} \\ 
         -\frac{\partial f_z(t)}{\partial t}\times[1-f^2_z(t)]^{-1/2}  \,\,\, {\rm if} \,\,\, t_{\rm ramp}< t\leq (t_\pi- t_{\rm ramp}) \\
          -\frac{\partial f_z(t_\pi)}{\partial t}\times[1-f^2_z(t_\pi)]^{-1/2} \times \sin^2({\pi (t-t_\pi)/2t_{\rm ramp}}) \,\,\, {\rm if} \,\,\, (t_\pi- t_{\rm ramp})<t\leq t_\pi
    \end{cases},
\end{equation}  
where $f_z(t)$ is defined in Eqs.~(8,9) of the main text and the pulse parameters can be found in Appendix~A. Also, we freely change a factor $\delta\Omega_{\pi}$ multiplying the Rabi frequency, i.e. $\Omega(t)\rightarrow(1+\delta\Omega_{\pi})\Omega(t)$, to ensure that $\int_0^{t_\pi}dt'\Omega(t')=\pi$. 

\begin{figure}
\centering
\includegraphics[width=1\linewidth]{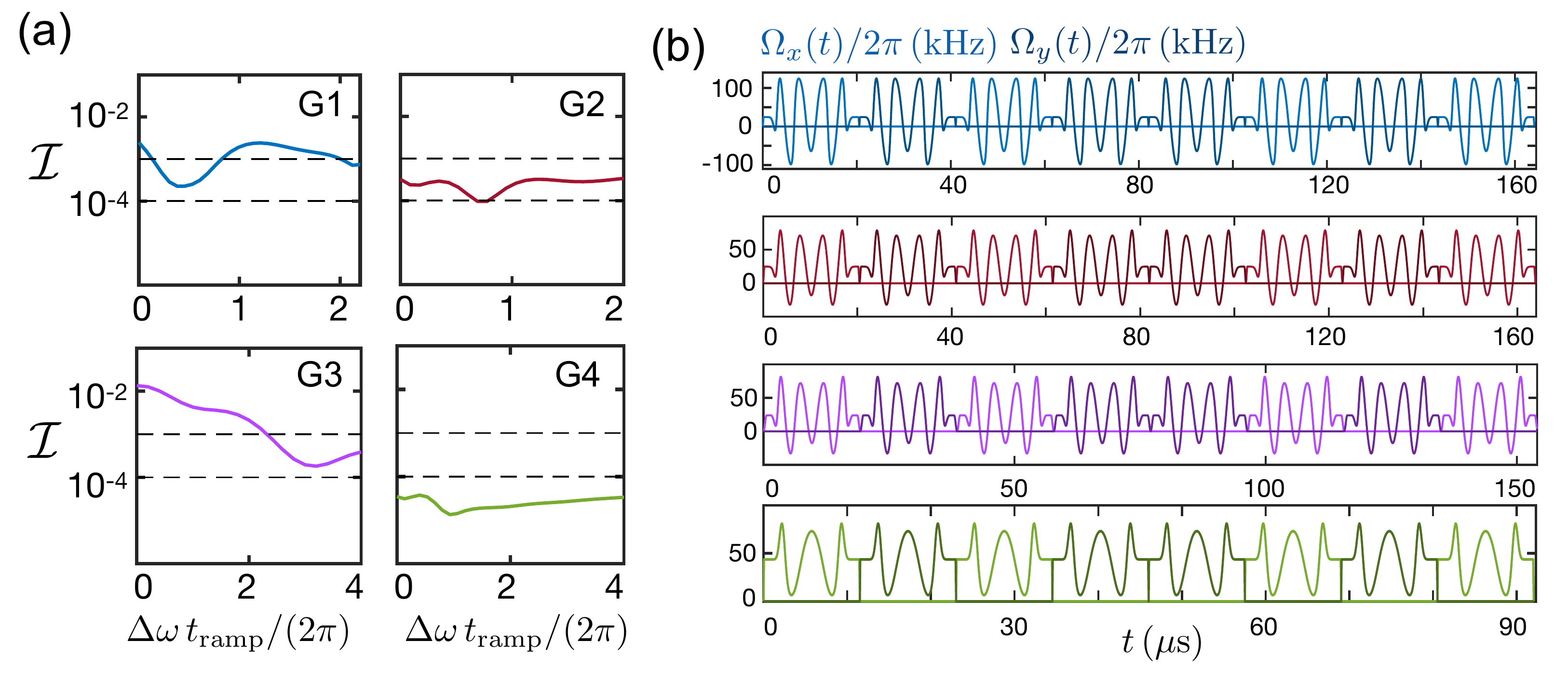}
\caption{ Optimisation of ramp duration (a) Bell state infidelity with respect to different ramp durations $t_{\rm ramp}$. For gates G1-4, we choose ramps with length $149, 295, 1260$ and $49$ ns, respectively. (b) The form of the Rabi frequencies $\Omega_x(t)$ (light line) and $\Omega_y(t)$ (dark line) is shown for gates from G1 (top figure) to G4 (bottom figure).}\label{fig:S3}
\end{figure}

We simulate the performance of the gate for different durations of the sin$^2$-shaped ramp $t_{\rm ramp}$. The obtained infidelities are shown in Fig.~\ref{fig:S3}(a). As it is shown, the achieved infidelity is in general better for the ramped case. The infidelities presented in table I of the main text are calculated using ramps of size $149, 295, 1260$ and $49$ ns, for gates G1-4, respectively. Because of the shift $\delta\Omega_{\pi}$, the maximum values of the Rabi frequencies $\Omega_{\rm pp}/2\pi$ change to $125.2, 79.05, 82.09$, and $81.14$ kHz, respectively. The Rabi frequencies $\Omega_x(t)$ and $\Omega_y(t)$ are shown in Fig.~\ref{fig:S3}(b), for gates G1-4, and for the first eight pulses.

\subsection{Heating of the bosonic mode}\label{subsec:heating}

To include the effect of motional heating we solve the master equation
\begin{equation}
\dot{\rho}=-i[H_s,\rho] +\mathcal{L}(\rho),
\end{equation}
where $\rho$ is the density matrix and the Lindblad superoperator is
\begin{eqnarray}
\mathcal{L}(\rho)= \frac{\Gamma}{2}(\bar{N}+1)(2a\rho a^\dagger - a^\dagger a \rho - \rho a^\dagger a) \nonumber + \frac{\Gamma}{2}\bar{N}(2 a^\dagger \rho a - a a^\dagger \rho - \rho a a^\dagger)
\end{eqnarray}
with $\bar{N}=[\exp{(\frac{\hbar\nu}{k_{\rm B}T})}-1]^{-1}$ and $T=300$~K. For $\bar{N}\gg1$, $\Gamma$ relates to the heating rate as $\dot{\bar{n}}=\Gamma\bar{N}$. For our simulations, we use a heating rate of $\dot{\bar{n}}=35$ ph/s for regime (i). We derived this from the rate $\dot{\bar{n}}_{\rm ref}=120$ ph/s given in Ref.~\cite{Barthel22_2} for $\nu_{\rm ref}=(2\pi)\times120$ kHz, and using the relation $\dot{\bar{n}}=\dot{\bar{n}}_{\rm ref} \times(\nu_{\rm ref}/\nu)^2$~\cite{Brownnutt15}. For regime (ii), we use a heating rate of  $\dot{\bar{n}}=100$ ph/s~\cite{Private}.

\section*{  Supplementary note 8: Robustness to mode decoherence}

Here, we explain how our gates can follow non-circular phase-space trajectories and obtain enhanced robustness against the mode decoherence. 
First, let us recall the propagator associated to an oscillator-mediated two-qubit phase gate,
\begin{equation}\label{Unitary2}
U(s)=\exp{[(\alpha(s)a^\dagger-\alpha^*(s)a)S_z]}\times \exp{[i\theta(s)S_z^2]}.
\end{equation}
where $\alpha(s)$ describes the phase-space trajectory, $s$ is proportional to time and
\begin{equation}
\theta(s)={\rm Im}\int_0^sds'\alpha^*(s')\dot{\alpha}(s') 
\end{equation}
is the two-qubit phase.
In the usual case, $\alpha(s)$ follows a circular trajectory $\alpha(s)\approx\alpha_{\rm circ}(s)=R_{\rm circ}(1-e^{i s})$, where $R_{\rm circ}$ is determined by the target two-qubit phase. 
At $s_{\rm g}=2\pi$, the mode-displacement is zero, $\alpha_{\rm circ}(s_{\rm g})=0$, and the phase is $\theta(s_{\rm g})=2\pi R_{\rm circ}^2$.
Typically, $\theta(s_{\rm g})=\pi/8$ is targeted, leading to $R_{\rm circ}=1/4$. 
That is, an operation where $\alpha(s)$ follows the phase-space path of a $1/4$ radius circumference equals a $\exp{[i\pi S_z^2/8]}$ phase-gate at $s_{\rm g}=2\pi$. 

To become more robust against mode decoherence, Haddadfarshi and Mintert~\cite{Haddadfarshi16_2} propose to follow trajectories related to cardioids instead. The simplest cardioid is parametrized by $\alpha_{\rm card}(s)=R_{\rm card}(e^{i s}-e^{i s/2})$, where $s=\{0,4\pi\}$.
At $s_{\rm g}=4\pi$, $\alpha_{\rm card}(s_{\rm g})=0$, and the phase is $\theta(s_{\rm g})=6\pi R_{\rm card}^2$, which, with the usual condition $\theta(s_{\rm g})=\pi/8$ leads to a cardioid with $R_{\rm card}={(4\sqrt{3})}^{-1}$. 

In our case, the two-qubit phase has a contribution from the circular trajectory, but also from the dispersive terms, 
\begin{equation}
\theta(t_{\rm g}) \approx \theta_{\rm circ}(t_{\rm g} )+ \theta_{\rm disp}(t_{\rm g} )= 2\pi R_{\rm circ}^2  +\frac{1}{2}\nu\eta^2J_k t_{\rm g },
\end{equation}
see Eq.~(6) of the main text. Because of that, the radius of the circular trajectory is smaller than $1/4$, and it is given by
\begin{equation}
R_{\rm circ}=\frac{1}{4}\Big\{\sqrt{1+4\eta^2J_k^2/f_k^2}+2\eta J_k/f_k\Big\}^{-1}.
\end{equation}
For the circular case, $R_{\rm circ}$ and $\xi_k$ can be determined exactly. However, we will describe a method to approximate its value since it will be useful for the non-circular path. For that, we approximate the final time of the gate as $t_{\rm g}=8N\tau_k\approx16\pi kN/\nu$, where $\nu$ is the mode frequency, $k$ is the selected harmonic and $N$ is the number of TQXY16 blocks in the gate. This approximation leads to the following equation 
\begin{equation}
2\pi R_{\rm circ}^2  +8\pi k\eta^2J_k N\approx \pi/8, 
\end{equation}
or
\begin{equation}
 R_{\rm circ}  \approx\frac{1}{4}\sqrt{1-64k\eta^2J_k N}.
\end{equation}
In the case of the cardioid, we have that 
\begin{equation}
\theta(t_{\rm g}) \approx \theta_{\rm card}(t_{\rm g} )+ \theta_{\rm disp}(t_{\rm g} )= 6\pi R_{\rm card}^2  +\frac{1}{2}\nu\eta^2J_k t_{\rm g },
\end{equation}
and, using the same approximation, we get
\begin{equation}\label{AppRadius}
 R_{\rm card}  \approx \frac{1}{4\sqrt{3}}\sqrt{1-64k\eta^2J_k N}.
\end{equation}
Notice that, once $\eta$, $k$ and $N$ are defined, Eq.~(\ref{AppRadius}) gives us a good guess about the value of the parameter $R_{\rm card}$ characterising the cardioid resulting in $\theta(t_{\rm g}) =\pi/8$.

For our $\alpha(t)$ to follow a trajectory different from that of a circle, we allow the variation of the pulse length $t_{\pi}=\tau_k/2$ (thus $\xi_k$) after every TQXY16 block. After the $j$-th TQXY16 block, at $t_j$, the value of $\alpha_j\equiv \alpha(t_j)$ is 
\begin{equation}\label{CardioidEq1}
\alpha_j=-i\eta \nu \int_0^{t_j} dt' f_z(t')e^{i\nu t'}=-i\eta \nu \int_{t_{j-1}}^{t_j} dt' f_z(t')e^{i\nu t'}+\alpha_{j-1}\approx-i\frac{\eta\nu f_k}{2}\int_{t_{j-1}}^{t_j} dt' e^{i\xi_k^j t'}+\alpha_{j-1}=\frac{\eta\nu f_k}{2\xi_k^j}(1- e^{i\xi_k^j (t_j-t_{j-1})})e^{i\xi_k^j t_{j-1}}+\alpha_{j-1}.
\end{equation}
where $t_0=0$ and $\alpha_0=0$. Using that the length of each block is $t_j-t_{j-1}=8\tau_k^j=16\pi k /(\nu-\xi_k)$, this can be rewritten as 
\begin{equation}\label{CardioidEq3}
\alpha_j=R_j\,\big(1-e^{i\varphi_j}\big)e^{i\phi_{j}}+\alpha_{j-1},
\end{equation}
where $R_j=\eta \nu f_k/2\xi_j$, $\varphi_j =16\pi k \xi_j/(\nu-\xi_j)$, and $\phi_j=\xi_j \times \sum_{j'=1}^{j-1} \varphi_{j'}/\xi_{j'}$. Also, $\phi_1=0$ and, for clarity, $\xi_j\equiv\xi_{k,j}$. Notice that varying $\xi_j$ changes the radius and the direction of the $j$-th trajectory and that concatenating trajectories with different radiuses $R_j$ will lead to a non-circular displacement in the phase space.

\begin{figure}
\centering
\includegraphics[width=1\linewidth]{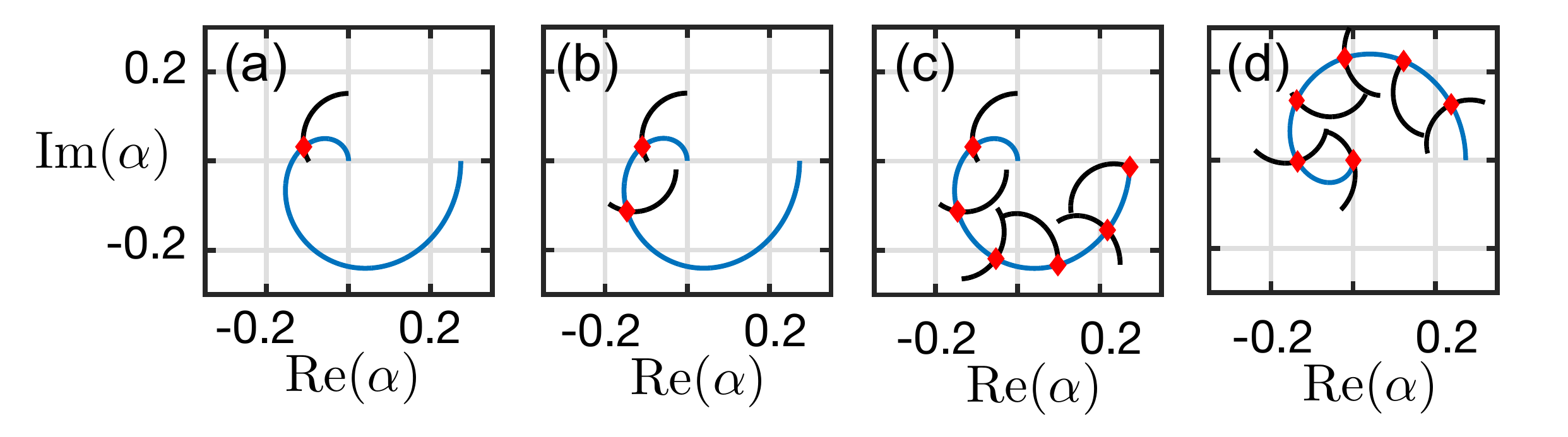}
\caption{Determination of $\xi_j$ for every TQXY16 block for G3. (a) From all $\xi_1=\{0,{\nu}/{8 k}\}$, we find the value of $\xi_1$ for which $\alpha_1$ crosses $\alpha_{\rm card}(s)$ for $s=\{0,2\pi\}$. In (b) and (c), the same is plotted but for $\xi_2$ and $\xi_6$, respectively. (d) shows the determination of the last six values for $\xi_j$, by matching the respective $\alpha_j$ with $\alpha_{\rm card}(s)$, for $s=\{2\pi,4\pi\}$. }\label{fig:S2}
\end{figure}

To find the values of $\xi_j$ leading to a path that approximates that defined by the cardioid characterised by $\alpha_{\rm card}(s)=R_{\rm card}(e^{i s}-e^{i s/2})$, where $R_{\rm card}$ is given in Eq.~(\ref{AppRadius}), we follow this procedure: (i) We choose $\eta$ and $k$ and, following the recipe in the main text, select a gate composed by $N$ TQXY16 blocks. Note that, when selecting $N$, the pulse parameters $b, c,$ and $d$ have also been determined. To building the cardioid trajectory, all blocks will stick to the same values of $b, c$ and $d$, while we will allow the variation of $t_{\pi}$ for every block. (ii) We propose a target cardioid with $R_{\rm card}=\frac{1}{4\sqrt{3}}\sqrt{1-64k\eta^2J_k \tilde{N}}$ where $\tilde{N}$ is an integer number approximately $122\%$ larger than than $N$. The reason behind this is that the perimeter of the cardioid is $122\%$ larger than that of the circle. (iii) Now, ee find all $\xi_j$ numerically. Once $\tilde{N}$ is defined, we obtain $\tilde{N}$ equations like Eq.~(\ref{CardioidEq3}), the first one with two unknowns, i.e. $\alpha_1$ and $\xi_1$, and the next ones depending also on $\alpha_{j-1}$ and previous values of $\xi_j$. We find the values of $\alpha_1$ and $\xi_1$ by numerically minimising $|\alpha_j(\xi'_j)-\alpha_{\rm card}(s)|$ for variables $\xi'_j=\{0,{\nu}/{8 k}\}$ and $s=\{0,2\pi\}$. In Fig.~\ref{fig:S2}(a), we illustrate this by showing $\alpha_{\rm card}(s)$ for $s=\{0,2\pi\}$ in blue, $\alpha_1(\xi'_1)$ for $\xi'_1=\{0,{\nu}/{8 k}\}$ in black. The value of $\alpha_1$ that crosses the cardioid $\alpha_{\rm card}(s)$ for $s=\{0,2\pi\}$ is shown with a red marker. After the value of $\alpha_1$ (thus $\xi_1$) has been numerically determined, we follow the same procedure for $\alpha_2$ and $\xi_2$, see Fig.~\ref{fig:S2}(b). We repeat this procedure until we find all $\alpha_j$ and $\xi_j$ for the first and second half of the cardioid with $s=\{0,2\pi\}$ and $s=\{2\pi,4\pi\}$ respectively, see Figs.~\ref{fig:S2}(c) and (d). The last step (iv) consist in checking the values of $\alpha_N$ and $\theta(t_{\rm g})$, which should give $0$ and $\pi/8$, respectively. If this is not case, we slightly change the values of $d$, $R_{\rm card}$ or $\tilde{N}$, and run the protocol described in (iii) until $\alpha_N\approx0$ and $\theta(t_{\rm g})\approx\pi/8$. Note that $\alpha_N$ and $\theta(t_{\rm g})$ can be calculated exactly using $\alpha(t_{\rm g})=-i \eta \nu \int_0^{t_{\rm g}}dt'f_z(t')\exp{(i\nu t')} $ and $\theta(t_{\rm g})=\int_{C}\alpha \, d\alpha$, where $f_z(t)$ is fully determined given all $\xi_k$.

\end{document}